\documentclass[useAMS,usenatbib]{mn2e}
\usepackage{epsf}
\usepackage{amssymb}
\usepackage[usenames]{color}

\def\plotone#1{\centering \leavevmode
\epsfxsize=\columnwidth \epsfbox{#1}}
\def\plottwo#1#2{\centering \leavevmode
\epsfxsize=.99\columnwidth \epsfbox{#1} \hfil
\epsfxsize=.99\columnwidth \epsfbox{#2}}
\def\plotwide#1{\centering \leavevmode
\epsfxsize=1.99\columnwidth \epsfbox{#1}}
\def\plotthree#1#2#3{\centering \leavevmode
\epsfxsize=.6\columnwidth \epsfbox{#1} \hfil
\epsfxsize=.6\columnwidth \epsfbox{#2} \hfil
\epsfxsize=.6\columnwidth \epsfbox{#3}
}
\def\kms{\,{\rm km~s}^{-1}}

\def\bnabla{\mbox{\boldmath $\nabla$}}
\newcommand{\be}{\begin{equation}}
\newcommand{\ee}{\end{equation}}
\newcommand{\half}{{\textstyle{1\over2}}}

\title[Isothermal potentials of elliptical galaxies]{Comparison of an
  approximately isothermal gravitational potentials of elliptical
  galaxies based on X-ray and optical data.}
\author[Churazov et al.]{E.~Churazov,$^{1,2}$ S.~Tremaine,$^{3}$
  W.~Forman,$^{4}$ O.~Gerhard,$^{5}$ P.~Das,$^{5}$ \newauthor
  A.~Vikhlinin,$^{4,2}$ C.~Jones,$^{4}$  H.~B\"ohringer,$^{5}$ K.~Gebhardt$^{6}$
\newauthor \\
$^1$ Max-Planck-Institut f\"ur Astrophysik, Karl-Schwarzschild-Strasse 1, 85741
Garching, Germany\\
$^2$ Space Research Institute (IKI), Profsoyuznaya 84/32, Moscow 117810, 
Russia\\
$^3$ Institute for Advanced Study, Einstein Dr., Princeton, NJ 08540, USA \\
$^4$ Harvard-Smithsonian Center for Astrophysics, 60 Garden St.,
Cambridge, MA 02138, USA \\
$^5$ MPI f\"{u}r Extraterrestrische Physik, P.O.\ Box 1603, 85740
Garching, Germany\\
$^6$ Department of Astronomy, University of Texas, Austin, TX 78712, USA \\
}

\begin{document}

\pagerange{\pageref{firstpage}--\pageref{lastpage}}
\pubyear{2009}

\maketitle

\label{firstpage}
\begin{abstract}
  We analyze six X-ray bright elliptical galaxies, observed with
  Chandra and XMM-Newton, and approximate their gravitational
  potentials by isothermal spheres 
  $\varphi=v_c^2\log r$ over a range of radii from $\sim$0.5 to $\sim
  25$ kpc. We then compare the circular speed $v_c$ derived from X-ray
  data with the estimators available from optical data. In particular
  we discuss two simple and robust procedures for evaluating the
  circular speed of the galaxy using the observed optical surface
  brightness and the line-of-sight velocity dispersion profiles. The
  best fitting relation between the circular speeds derived from
  optical observations of stars and X-ray observations of hot gas is
  $v_{c,opt} \simeq \eta~v_{c,X}$, where $\eta=1.10-1.15$ (depending
  on the method), suggesting, albeit with large statistical and systematic
  uncertainties, that non-thermal pressure on average contributes
  $\sim$20-30\% of the gas thermal pressure.
\end{abstract}


%

\sloppypar

\section{Introduction}
In spiral galaxies, disk rotation curves offer an accurate and robust
way of measuring total gravitational potentials to distances as
large as 10--30 kpc. To a first approximation the rotation curves are
flat over a broad range of radii, suggesting an isothermal
(logarithmic) potential characterized by $\varphi(r)=v_c^2\log r$. In
early-type galaxies, measuring the gravitational potential is much more
difficult since there are no tracers such as cold gas or disk stars
on orbits of known shape.  Several methods have been used to measure
the potentials of elliptical galaxies including detailed modeling of
stellar orbits \citep{2000A&AS..144...53K,2006MNRAS.366.1126C,2007MNRAS.382..657T,2009ApJ...695.1577G},
tracers such as globular clusters, planetary nebulae, and satellite
galaxies
\citep{2001ApJ...553..722R,2009MNRAS.394.1249C,2009ApJ...690.1488K},
strong and weak lensing of quasars and background galaxies
 \citep{2006ApJ...649..599K,2006MNRAS.368..715M,2007ApJ...667..176G,mvk08} 
and (for the most massive galaxies)
modeling the hydrostatic atmospheres of X-ray emitting gas
\citep{1978ApJ...219..413M,1985ApJ...293..102F,2006ApJ...636..698F,2006ApJ...646..899H,2008MNRAS.388.1062C}.
Recent studies using stellar kinematics and lensing suggest that the
potentials of early-type galaxies are approximately isothermal
\citep{2001AJ....121.1936G,2006ApJ...640..662T,2007ApJ...667..176G},
similar to disk galaxies. Consistency of the mass profile with
  $M(r)\propto r$  (implying an isothermal potential) was
  also suggested for several elliptical galaxies based on the analysis
  of X-Ray data \citep[e.g.,][]{1985ApJ...296..447T,1994ApJ...427...86B,1995ApJ...441..182K,1995MNRAS.274.1093N,1998MNRAS.298..811B,2006ApJ...636..698F}. 
Here we present independent evidence that
the potentials of bright elliptical galaxies are close to isothermal,
from X-ray observations with Chandra and XMM-Newton. If the potential of a typical elliptical galaxy is
  indeed not far from being isothermal (for a range of radii) then it can be
  characterized in that radius range with a single number -- the circular speed
$v_c$. This makes the comparison of potentials derived
from X-ray and optical data especially simple since it does not require
a point-by-point comparison and for a pure isothermal (logarithmic)
potential the results are not sensitive to the range of radii used for
evaluation of $v_c$. While deviations from isothermality
are certainly present at some level, we believe this approach is useful. In
this paper we develop simplified methods for characterizing the X-ray and optical
data and comparing the results to place constraints on the non-thermal
pressure in the hot gas in elliptical galaxies.

By construction our method is quick and approximate and is not
intended to  replace a careful and comprehensive analysis of
individual objects. It might be useful, for example, in a larger 
sample, when detailed modeling is not practical due to noisy or missing 
data. We illustrate the method on a small and rather
arbitrarily selected sample of X-ray bright elliptical galaxies.
 
The structure of this paper is as follows: in \S\ref{sec:sample} we describe
our sample of galaxies and how we derive the gravitational potential from
X-ray observations, and in \S\ref{sec:robust} we discuss
methods for determining the potential from optical observations of stellar
velocity dispersions. The implications of our results for the mass
distribution and non-thermal pressure contribution in these galaxies are discussed in
\S\ref{sec:discussion} and \S\ref{sec:conc} contains conclusions. 

\section{The sample and X-ray analysis}
\label{sec:sample}
For our analysis we selected six nearby (distance less than $\sim$30 Mpc, see
Table \ref{tab:sample}), X-ray bright galaxies, which were all well
observed with Chandra and XMM-Newton. All galaxies in
the sample are very bright and dominate (in terms of mass or
potential) their environment up to at least several effective
radii. This (at least partly) justifies the standard practice of deriving
mass/potential 
profiles from X-ray data using the assumption that the X-ray
emitting gas forms a hydrostatic atmosphere.

\begin{table*}
\centering
\caption{Sample of elliptical galaxies. \label{tab:sample} The columns
are: (1) - common name of the galaxy; (2) - redshift from the NASA/IPAC
Extragalactic Database; (3) - adopted distance; (4) - hydrogen column
density from \protect\cite{2005A&A...440..775K}; (5) effective radius (eq.\ \ref{eq:reff}), (6)
S\'{e}rsic index, (7) - central 
velocity dispersion standardized to $\sim0.6$ kpc aperture from Hyperleda, 
except for NGC 4472, taken from B94, 
(8) - line-of-sight velocity dispersion
at the ``sweet spot'' $R_s$; (9) - circular speed estimated from the central velocity
dispersion as $v_{c,c}=\sqrt{2}\sigma_{c}$, 
(10) - circular speed $v_{c,s}$ according to \S\ref{sec:methoda},
(11) - circular speed $v_{c,l}$ according
to \S\ref{sec:methodb}. References for the effective radius,
S\'{e}rsic index and stellar kinematics: [D94] - \protect\cite{1994MNRAS.271..523D}; [C93]
- \protect\cite{1993MNRAS.265.1013C}; [S08] - \protect\cite{2008MNRAS.385..667S};
[M05] - \protect\cite{2005AJ....130.1502M}; [K00] -
\protect\cite{2000A&AS..144...53K}; 
[S00] - \protect\cite{2000AJ....119..153S};
[B94] - \protect\cite{1994MNRAS.269..785B}; [G09] -
\protect\cite{2009ApJ...700.1690G,gebhardt2009b}, [D01] -
\protect\cite{2001ApJ...546..903D}. 
Distances are from \protect\cite{2001ApJ...546..681T}. Note that 
distances are not explicitly used in the subsequent analysis.}
\begin{tabular}{lcrllllllll}
\hline
Name   & $z$ &$D$, Mpc & $ N_H$& $R_e$, arcsec & S\'{e}rsic index,
$n$ & $\sigma_{c}$&  $\sigma(R_s)$ &$v_{c,c}$ &$v_{c,s}$&$v_{c,l}$ \\
(1)   & (2) & (3) & (4) & (5) & (6) & (7) &  (8) & (9) & (10) & (11) \\
\hline
NGC1399        &0.00475 & 20.0 &$1.5~10^{20}$& 117 & 12.24    [D94] & 341 & 242 [S00] &482& 412 & 394\\
NGC1407        &0.00593 & 28.8 &$5.4~10^{20}$& 70  & 8.35$~$  [S08] & 272 & 256 [S08] &385& 435 & 408\\
NGC4472 (M49)   &0.00333 & 16.3 &$1.5~10^{20}$& 257 & 6.27$~$ [C93] & 320 & 289 [B94] &452& 492 & 445\\
NGC4486 (M87)  &0.00436 & 16.1 &$1.9~10^{20}$& 145 & 6.51$~$ [C93]  & 336 & 312 [G09] &475& 530 & 536\\
NGC4649 (M60)  &0.00373 & 16.8 &$2.0~10^{20}$& 118 & 5.84$~$ [C93]  & 336 & 244 [D01] &475& 414 & 436\\
NGC5846        &0.00572 & 24.9 &$4.3~10^{20}$& 79  & 3.95$~$ [M05]  & 241 & 215 [K00] &341& 366 & 338\\
\hline
\end{tabular}
\end{table*}

For the analysis we used publicly available Chandra and XMM-Newton
data. Combining the data from these two instruments 
provides a cross-check of the results and also gives better
constraints on the innermost and outermost regions, thanks to the superb
angular resolution of Chandra and the large field of view of XMM-Newton,
respectively.

For Chandra the data were prepared following the procedure described
in \cite{2005ApJ...628..655V}. This includes filtering of high
background periods and application of the latest calibration
corrections to the detected X-ray photons, and determination of the
background intensity in each observation.

For XMM-Newton the data were prepared by removing background flares
using the light curve of the detected events above 10 keV and
re-normalizing the ``blank fields'' background to match the observed
count rate in the 11-12 keV band.  In the subsequent analysis we use
the data from the EPIC/MOS detector only. 

The analysis of the X-ray data is based on a non-parametric deprojection
procedure, described in \citet[C08
hereafter]{2003ApJ...590..225C,2008MNRAS.388.1062C}. In brief, the observed
X-ray spectra in concentric annuli are 
modeled as a linear combination of spectra in spherical shells; the two
sequences of spectra are related by a matrix describing the projection of the
shells
into annuli.  To account for the projected
contribution of the emission from shells at large distances
from the center (i.e., at distances larger than the radial size
$r_{max}$ of the region well covered by actual observations) one has
to make an explicit assumption about the behavior of the gas
density/temperature profile at large radii. We assume that at all energies the
gas volume emissivity at radii beyond $r_{max}$ declines as a power law with
radius. The slope of this power law is estimated based on the
observed surface-brightness profile within the range of radii covered
by observational data. Since we assume that the same power law shape
is applicable to all energy bands, 
effectively this assumption
implies constant spectral shape and therefore the
isothermality of the gas outside $r_{max}$ (here ``isothermal'' means
that for $r\ge r_{max}$ the
gas temperature is independent of radius, not that the gravitational potential
is logarithmic). The contribution of these
layers is added to the projection matrix with the normalization as an
additional free parameter.  While the limitations of this approach are
obvious, the contribution of these outer shells is usually
important only in the few outermost radial bins inside $r_{max}$, especially when
the surface-brightness profile is steep. The final projection matrix
is inverted and the shells' spectra are explicitly calculated by
applying this inverted matrix to the data in narrow energy channels.

The resulting spectra are approximated in XSPEC \citep{1996ASPC..101...17A}
with the APEC one-temperature optically thin plasma emission model
\citep{2001ApJ...556L..91S}.  The redshift $z$ (from the NASA/IPAC
Extragalactic Database -- NED) and the line-of-sight column density of neutral
hydrogen $N_H$ (based on Kalberla et al. 2005) have been fixed at the values
given in Table \ref{tab:sample}. For each shell we determine the emission
measure (and therefore gas density) and the gas temperature. These quantities
are needed to evaluate the gravitational potential through the hydrostatic
equilibrium equation.  For cool (sub-keV) temperatures and approximately solar
abundance of heavy elements, line emission provides a substantial fraction of
the 0.5-2 keV flux. With the Chandra and XMM-Newton spectral resolution the
contributions of continuum and lines are difficult to disentangle. As a result
the emission measure and abundance are anti-correlated, which can lead to
large scatter in the best-fit emission measures. As an interim (not entirely
satisfactory) solution, we fix the abundance at 0.5 solar for all shells, using
the default XSPEC abundance table of \cite{1989GeCoA..53..197A}. We return to
this issue in \S\ref{sec:metals}.

\begin{table*}
\centering
\caption{Best fitting $v_{c,X}$ derived from Chandra and XMM-Newton
  data on the potential profiles approximated with a logarithmic
  law $\varphi(r)=v_c^2 \log r + b$. \label{tab:sigmax} The quoted uncertainties are pure
  statistical errors, determined from a Monte Carlo procedure. The last
  column is the average of the results from Chandra and XMM-Newton in the
  preceding two columns.}
\begin{tabular}{llllll}
\hline
Name   &$r_{1}(')$ & $r_{2}(')$& $v_{c, \rm Chandra}\kms$&  $v_{c,\rm XMM}\kms$ &  $v_{c,X}\kms$\\
\hline
NGC1399       & 0.1 & 5.0 & $403\pm 1.8$ & $395\pm 2.1$ & 399 \\
NGC1407       & 0.1 & 2.0 & $368\pm 9.8$ & $356\pm 9.8$ & 362 \\
NGC4472 (M49) & 0.1 & 5.0 & $372\pm 4.0$ & $367\pm 2.0$ & 370 \\
NGC4486 (M87) & 0.1 & 5.0 & $448\pm 1.3$ & $437\pm 1.6$ & 443 \\
NGC4649 (M60) & 0.1 & 5.0 & $417\pm 2.4$ & $422\pm 1.8$ & 420 \\
NGC5846       & 0.1 & 5.0 & $335\pm 3.1$ & $331\pm 5.2$ & 333 \\
\hline
\end{tabular}
\end{table*}

\begin{figure*}
\plotwide{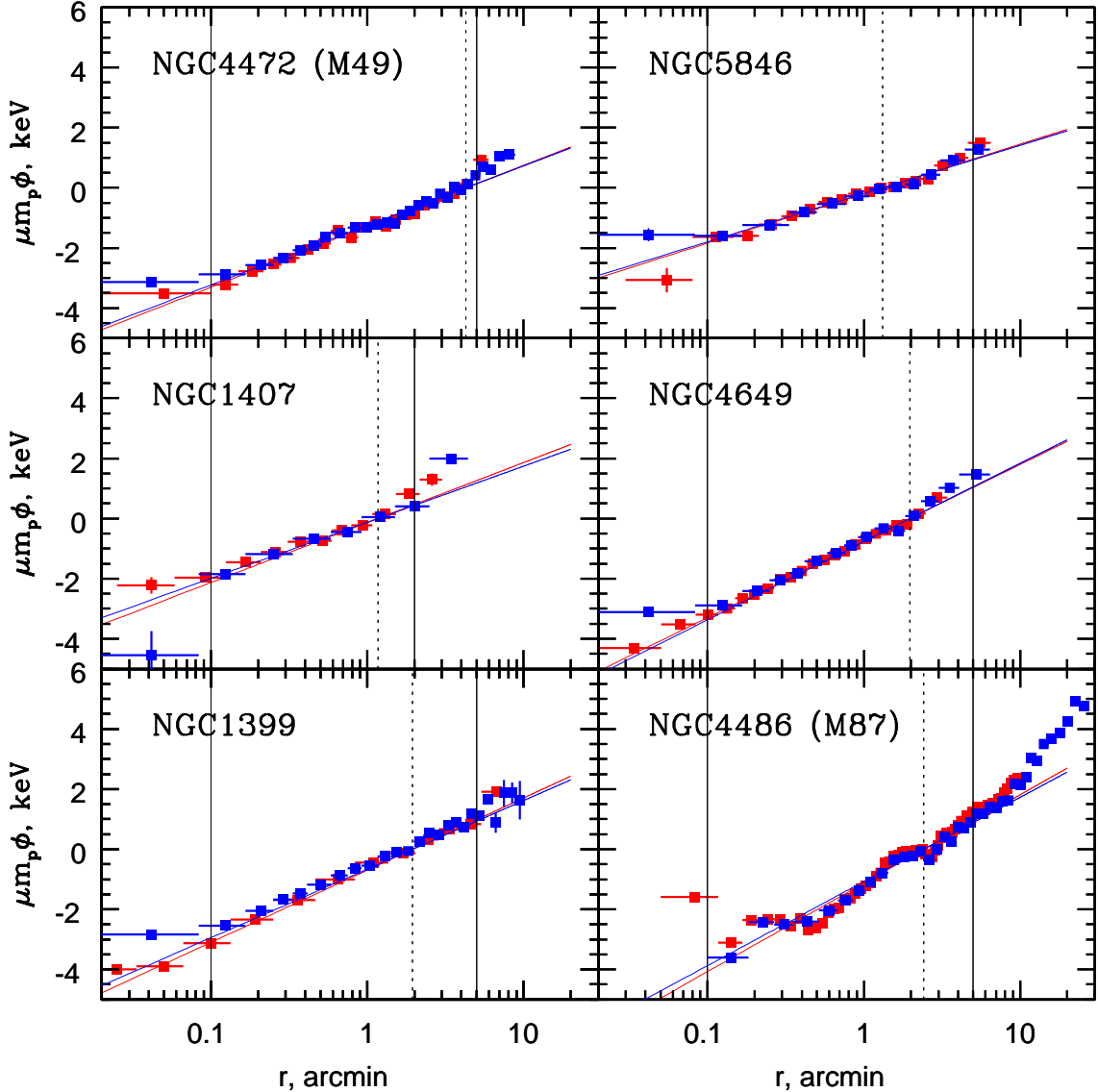}
\caption{Gravitational potentials derived from Chandra (red) and
  XMM-Newton/MOS (blue) observations of each galaxy. Potentials are
  normalized to zero at $R_e$.  
Vertical lines mark the range of radii used to
approximate the data by a $v_c^2 \log r+b$ law, and the thin solid lines are
the best-fit approximations of the Chandra and XMM-Newton data to this
law. The vertical dotted lines mark the effective radius.
\label{fig:pind}
}
\end{figure*}

With known gas density $n$ and temperature $T$ in each shell we can
use the hydrostatic equilibrium equation to evaluate the gravitational
potential $\varphi$:
\begin{eqnarray}
\frac{1}{\rho}~\frac{dP}{dr}=-\frac{d\varphi}{dr}
\label{eq:hyd}
\end{eqnarray}
where $\rho=\mu m_p n$ is the gas density, $P=nkT$ is the pressure,
$\mu$ is the mean atomic weight of the gas ($\mu=0.61$ assumed
throughout the paper), $m_p$ is the proton mass and $k$ is the
Boltzmann constant. Integrating the above equation one gets an
expression for the gravitational potential through the observables $n$
and $T$:\footnote{Throughout this paper, ``$\log$'' denotes natural
  logarithm.} 
\be
\varphi=-\frac{k}{\mu
m_p}\left[ \int{ T\frac{d\log n}{dr} dr} +T \right]+C, 
\label{eq:pot}
\ee
where $C$ is an arbitrary constant. We choose the constant $C$ such that
$\varphi(R_e)=0$, where $R_e$ is the optical effective radius (see eq.\
\ref{eq:reff} and Table \ref{tab:sample}). The resulting potentials are shown
in Figure \ref{fig:pind} with red (Chandra) and XMM-Newton (blue) points. The
error bars were evaluated by a Monte Carlo procedure, starting from the measured values of
$n$ and $T$ in each shell, adding ``noise'' to the data points and re-deriving
the potential via equation (\ref{eq:pot}) (see C08 for a discussion of the
limitations of this procedure).

 As discussed in C08 the expression (\ref{eq:pot}) for the
  potential can be evaluated directly from deprojected data - density
  and temperature profiles. Since the logarithmic derivative of the
  density $\frac{d\log n}{dr}$ is under the integral the enhancement
  of errors due to differentiation is not a major issue and evaluation
  of eq. (\ref{eq:pot}) does not require any parametric description of
  the density and/or temperature profiles. This removes the ambiguity
  in the choice of a functional form for these profiles. Since the
  potential is calculated directly via eq. (\ref{eq:pot}) no apriori
  parametrization of the potential (or mass) is required.  If
  information on the gravitational potential is available from other
  data (e.g. from optical observations) the comparison of the
  potential profiles can be done directly with the results of
  eq. (\ref{eq:pot}). In C08 the relation between potential profiles
  derived from X-ray and optical data
  was written as $\varphi_X=a \varphi_{opt}+b$ and the value of $a$
  was evaluated. In this case any transformation of the potential
  profile to the circular speed (or equivalently to the mass profile)
  which involves differentiation of the potential profile is not
  needed and should be avoided. 

  Thus, the application of eq. (\ref{eq:pot}) to the X-ray data is
  essentially free from any parametrization and does not rely on
  direct differentiation of the data. This is the main difference with
  most of the other techniques which either use parametrization of the
  density/temperature profiles or make an assumption on the form of
  the mass profile \cite[see e.g.][for various techniques of the mass
    profile
    reconstruction]{1978ApJ...219..413M,1981ApJ...248...47F,1985ApJ...293..102F,1995MNRAS.274.1093N,2006ApJ...646..899H,2006ApJ...636..698F}. The
  most close approach to ours is that of \citet{2006ApJ...646..899H},
  where the temperature and mass profiles are parametrized, the
  hydrostatic equilibrium equation is solved to find the density and
  the best-fitting parameters are iteratively found. As mentioned
  above in eq. (\ref{eq:pot}) all observables are on the r.h.s and one
  can reconstruct the potential directly and
  defer the parametrization 
  (if needed) to a final step of manipulations with reconstructed potential.

  We are now considering the case when limited
  information on the potential from optical data is available. For
  instance, consider a situation when only central velocity dispersion
  of the galaxy is known (see discussion in \S\ref{sec:robust}). In
  this case to compare X-ray and optical potentials a parametrization
  of the potential derived from eq. (\ref{eq:pot}) is needed. An
  especially simple and straightforward parametrization is possible if
  the potential is isothermal $\varphi=v^2_c \log r +b$, since in this
  case a single number -- circular speed $v_c$ can characterize the
  potential. In fact, recent studies using stellar kinematics and
  lensing do suggest that the potentials of early-type galaxies are
  approximately isothermal
  \citep{2001AJ....121.1936G,2006ApJ...640..662T,2007ApJ...667..176G},
  similar to disk galaxies. Consistency of the mass profile with
  $M(r)\propto r$ (implying an isothermal potential) was also
  suggested for several elliptical galaxies based on the analysis of
  X-Ray data
  \citep[e.g.,][]{1985ApJ...296..447T,1994ApJ...427...86B,1995ApJ...441..182K,1995MNRAS.274.1093N,1998MNRAS.298..811B,2006ApJ...636..698F}. Below
  we confirm an approximate isothermality of the potentials using the
  results of application of the non-parametric method
  (eq. \ref{eq:pot}) and determine the best-fitting value of $v_c$.

The axes in Figure \ref{fig:pind} are log-linear and therefore an isothermal
(logarithmic) potential should look like a straight line $v^2_c \log r +b$. To
first order this is true, although there are statistically significant
deviations, particularly at the innermost and outermost radii, which we
discuss below. The agreement of Chandra and
XMM-Newton data is good, except for the inner region where the better spatial
resolution of Chandra is important. We choose to ignore the data inside the $r_1=
0.1'$ circle where this effect is apparent.  We also introduce a cutoff at
large radii (typically $r_2\approx 5'$), where the results are sensitive to
the assumed extrapolation of the emissivity profile. The actual values of
$r_1$ and $r_2$ used in the analysis are given in Table \ref{tab:sigmax} and
are shown in Figure \ref{fig:pind} as a pair of vertical lines. Between $r_1$
and $r_2$ the potential was approximated with a logarithmic function
$v_c^2 \log r + b$ with $v_c$ and $b$ being free parameters. Best
fitting values were found by minimizing the root-mean-square deviation between
the model and observed potential profiles. The resulting values of $v_{c,X}$
for Chandra and XMM-Newton data are given in Table \ref{tab:sigmax} as
$v_{c,\rm Chandra}$ and $v_{c,\rm XMM}$. The statistical uncertainties
determined from the Monte Carlo procedure are also given in Table
\ref{tab:sigmax}. In general there is good agreement between the values
obtained by the two instruments, confirming that the uncertainties introduced
by statistical errors and cross-calibration uncertainties between the two
instruments (including background subtraction procedures) are small. One can expect however that the real uncertainties are
dominated by systematic errors arising from our model assumptions (e.g., the
assumption of spherical symmetry used in the deprojection analysis), which
affect both datasets in similar ways. For subsequent analysis we use the
average of the results from the two instruments, 
$v_{c,X}\equiv (v_{c,\rm Chandra}+v_{c,\rm XMM})/2$ (last column in Table
\ref{tab:sigmax}). 

The deviations of the potential from the isothermal shape
  $\varphi=a \log r +b$ can be studied by assuming a functional form
  $\varphi=ar^\alpha+b$ and looking for the best-fitting value of
  $\alpha$. For the five galaxies with
  $r_1= 0.1'$ and $r_2=5'$ the best-fitting value of $\alpha$
  varies from $0.03$ for NGC1399 to $0.16$ for NGC4472 (the mean value
  of $\alpha$ is 0.11). For NGC1407
  ($r_1=0.1'$ and $r_2=2'$) the value of $\alpha$ is 
  $\sim 0.5$. This suggests that on average the profiles are slightly
  concave, i.e. circular speed slightly increases at large radii,
  which can be seen in Fig.\ \ref{fig:pind}. This is consistent with the
  fact that these massive ellipticals are sitting at the centers of
  more massive group/cluster size halos which dominate at large
  radii. Given the mean value of $\alpha=0.11$ one can estimate that
  changing both $r_1$ and $r_2$ by a factor of two would on average change the
  estimate of $v_c$ by a factor of only $\displaystyle
  2^{\alpha/2}\sim4$\%. 

We explicitly tested the sensitivity of $v_c$ to the values of $r_1$ and
  $r_2$ for all objects in the sample by increasing the lower boundary
  by a factor of two (i.e. $r_1=0.2'$) and recalculating $v_c$. The
  results of this test are given in Table  \ref{tab:errx} (column
  $\Delta_{\rm r_1\times 2}$), where
\begin{eqnarray} 
\Delta_{\rm r_1\times 2}=\frac{v_{c,{\rm r_1\times 2}}-v_{c,X}}{v_{c,X}},
\label{eq:delta}
\end{eqnarray}
and $v_{c,X}$ is the circular speed from Table \ref{tab:sigmax}, and $v_{c,{\rm r_1\times 2}}$ is calculated as a mean value of
circular speeds measured using Chandra and XMM-Newton for increased
$r_1$ (similarly to $v_{c,X}$ in Table \ref{tab:sigmax}). In a
separated test we decreased the value of $r_2$ by a factor of two
(i.e. $r_2=2.5'$ for all objects except for NGC1407, where $r_2=1'$)
and again recalculated $v_c$ (see Table  \ref{tab:errx}, column $\Delta_{\rm
  r_2/2}$). 
From Table \ref{tab:errx} it follows that factor of 2 changes in
either $r_1$ or $r_2$ causes few per cent changes in $v_c$.

\subsection{Flat abundance profile}

\label{sec:metals}

We now illustrate the impact of our assumption of a flat abundance
profile, using NGC1399 as an example. 

As is obvious from equation (\ref{eq:pot}), the absolute normalization
of the gas density $n$ does not affect the calculations of the
potential. From this point of view it is not the particular value of
the heavy-element abundance in the spectral models, but rather the
radial variation of the abundance that affects the derived potential profile. In
elliptical galaxies one can expect an increase of metal abundance
towards the center of the galaxy. This is usually true for a range of
radii except for the very center, where a ``dip'' in the metal
abundance is often seen when fitting the data with a single
temperature plasma emission model \citep[e.g.,][]{2002A&A...386...77M}.

As mentioned above, measuring the metal abundance from X-ray spectra in cool
systems is difficult because of the ambiguity of separating line from
continuum emission with the limited energy resolution of X-ray CCDs. For a
multi-temperature plasma this separation is even more complicated because
fitting the emission with one-temperature models leads to a biased estimate of
the abundance \citep[e.g.,][]{2000ApJ...539..172B}. In deprojected spectra,
such as we use here, the bias arising from a multi-temperature plasma is
reduced because the superposition of cooler and hotter emission coming
from different radii is removed (provided the object is spherically
symmetric), although if the plasma is intrinsically multi-temperature then the
problem remains. On the other hand the signal-to-noise of the deprojected
spectra is much lower than for the projected spectra. It is therefore
desirable to keep the number of free parameters in the fit as small as
possible.

\begin{figure*}
\plottwo{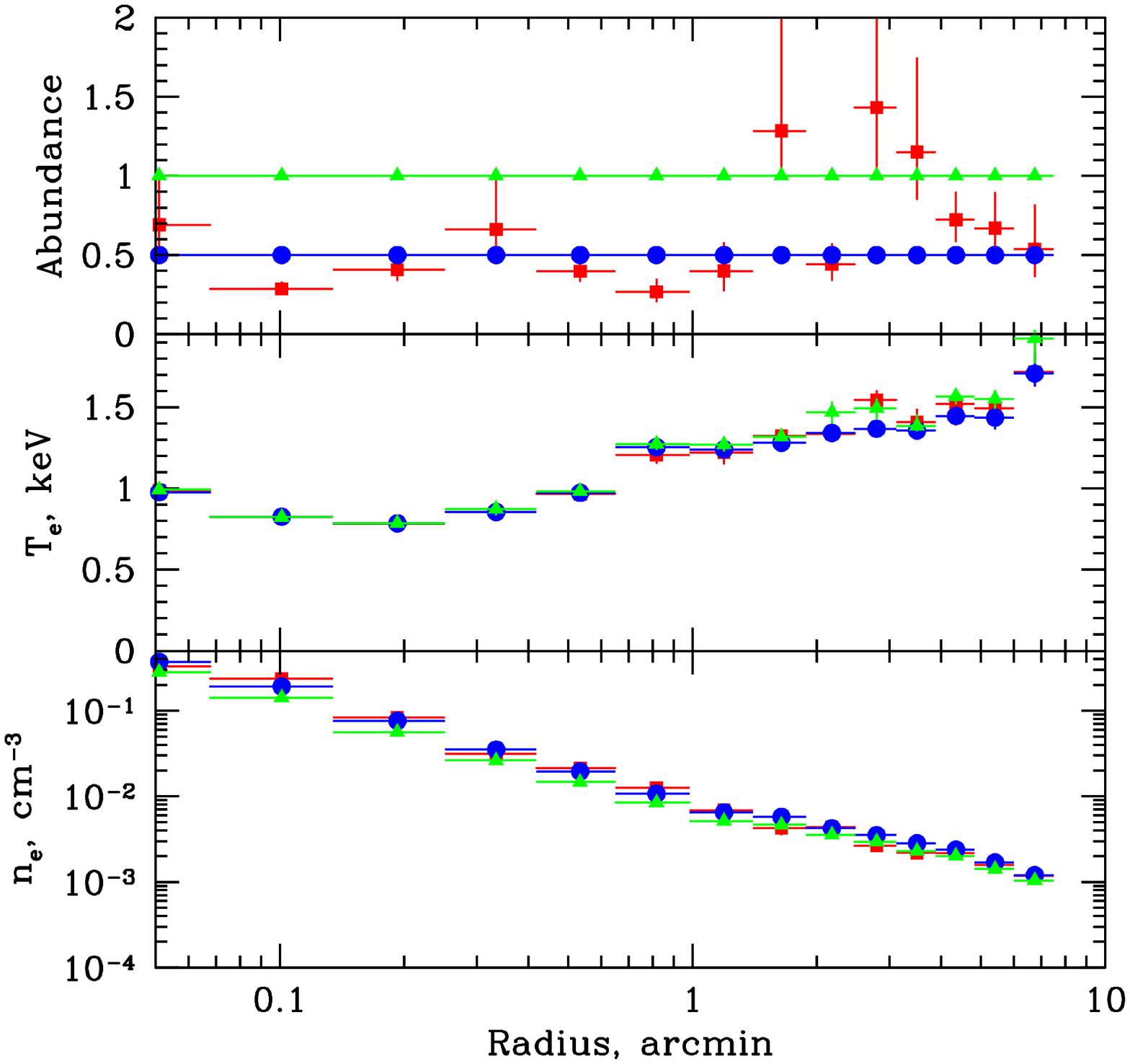}{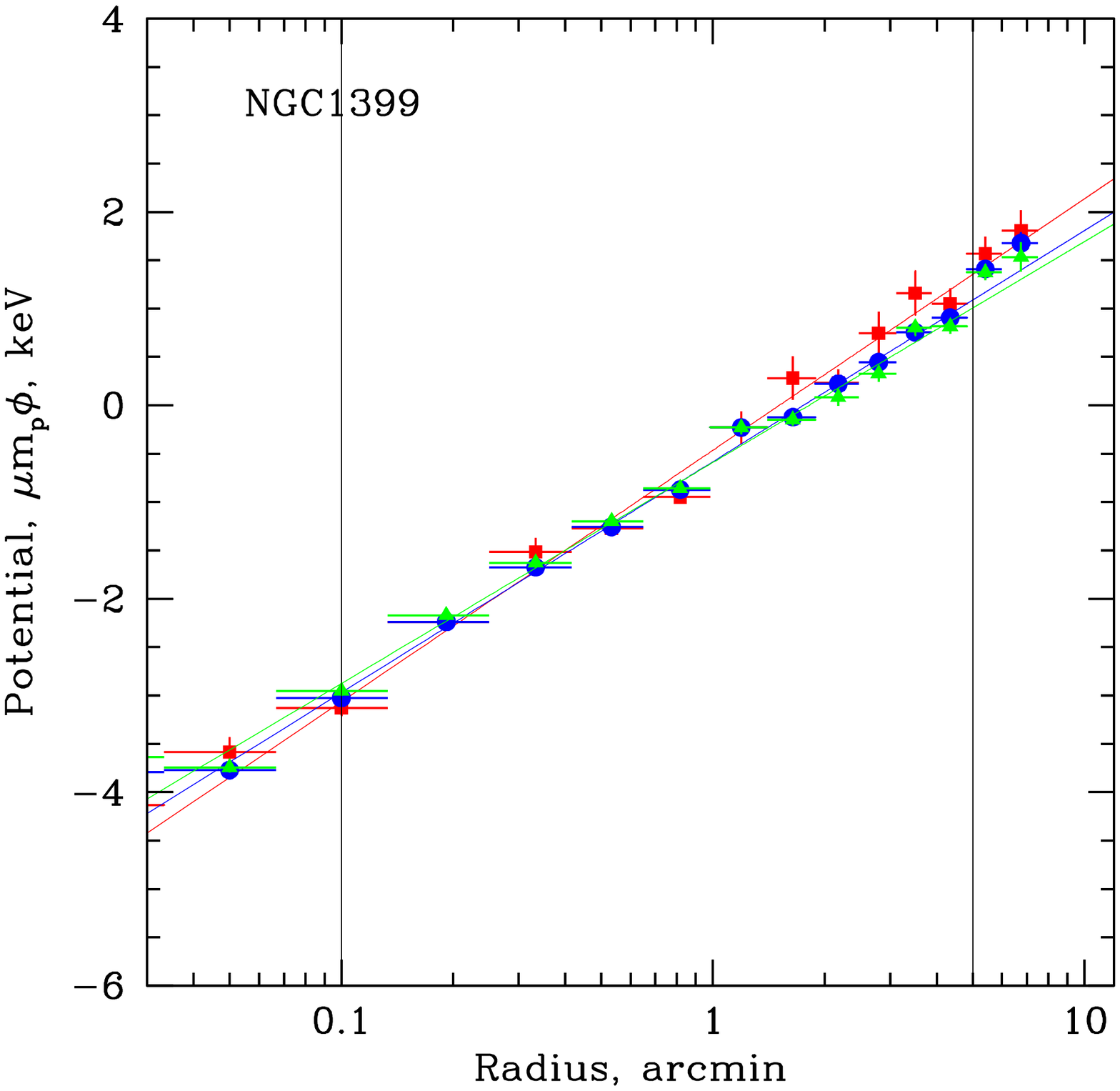}
\caption{{Left:} Radial profiles of deprojected gas parameters (abundance,
  temperature, electron density) in
  NGC1399. The parameters were obtained using a single-temperature
  APEC model (with fixed low-energy absorption and redshift) fit to
  the deprojected Chandra spectra from a set of spherical shells. For the blue
  and green models (circles and triangles) the abundance was fixed at 0.5
  and 1 times solar, respectively, while for the red model (squares) the
  abundance was a free function of the radius. 
{Right:} Potential profiles corresponding to the spectral models
shown in the left panel. Straight lines show best-fitting
approximations $\varphi=v_c^2\log r + b$. 
\label{fig:abund}
}
\end{figure*}

As an example we show in Figure \ref{fig:abund} (left) the gas parameters
(electron density, temperature and abundance of heavy metals) as a
function of radius for NGC1399. In two models (blue circles and green triangles)
the abundance was fixed at 0.5 and 1 times solar, respectively. In the third 
model (red squares) the abundance was a free parameter. One can see that in the third model 
the errors on the abundance are substantial and, as expected, the
abundance and the gas density are anti-correlated. The potential
profiles corresponding to these spectral models are shown in the
right panel of Figure \ref{fig:abund}. Clearly there is a significant change
in the potential curves (curves are normalized to zero potential at
$1'.5$). Formally calculated values of $v_{c}$ for the three
spectral models are:
\begin{eqnarray}
\begin{array}{ll}
v_{c} &=421~\kms; \hbox{ free abundance} \\
v_{c} &=403~\kms; \hbox{ abundance$\,=0.5\ \times$ solar} \\
v_{c} &=396~\kms; \hbox{ abundance$\,=1\ \times$ solar}. \\
\label{eq:abund} 
\end{array}
\end{eqnarray}
Thus, there is a substantial, but not dramatic effect of the assumed abundance
profile on the derived logarithmic slope of the potential profile. Given that
the limited statistics in our deprojected spectra do not allow for a robust
abundance determination for all galaxies in the sample, we decided to keep the
assumption of a flat abundance profile so that we could analyze all objects in
a uniform way. It is likely that this approximation introduces errors in the
best-fitting value of $v_{c}$ of roughly 5-15$\kms$. If all
galaxies in our sample have a similar abundance profile to the one derived in
NGC1399, one can expect the values of $v_{c}$ obtained under the
assumption of a flat profile with $0.5\ \times$ solar metallicity to be
biased low by ~3-4\%. 

Relative changes in the circular speed when abundance of heavy
  elements is a free parameter (constrained to be in the range from 0
  to 2) for all objects in the sample are calculated in Table
  \ref{tab:errx} (column $\Delta_{\rm abund}$):
\begin{eqnarray} 
\Delta_{\rm abund}=\frac{v_{c,{\rm
      free~abundance}}-v_{c,{X}}}{v_{c,{X}}},
\label{eq:delta1}
\end{eqnarray} 
where $v_{c,{\rm free~abundance}}$ is calculated as a mean value of
circular speeds measured using Chandra and XMM-Newton, similarly to
$v_{c,X}$ (see \S\ref{sec:sample}).

Yet another parameter related to the chemical composition of the hot gas that
is important for the evaluation of the gravitational potential is the mean
atomic weight $\mu$ (see eq.\ \ref{eq:pot}). For the solar photospheric
abundance table of \cite{1989GeCoA..53..197A} $\mu=0.614$ (assuming fully
ionized plasma). If we lower the abundance of heavy elements to 0.5
solar, $\mu$ changes
 to 0.610, i.e. by 0.7\%. Changing the helium abundance would, of course,
have a much stronger impact on $\mu$. For instance, doubling the abundance of
helium relative to the solar photospheric mix would increase $\mu$ to 0.695,
i.e., by 13\%. One mechanism that could lead to helium enrichment beyond solar
abundance is gravitational sedimentation of heavy elements
\citep[e.g.,][]{1984SvAL...10..137G,2003MNRAS.342L...5C,2004MNRAS.349L..13C,2006MNRAS.369L..42E}.
If sedimentation is indeed important for the interstellar gas in the cores of
elliptical galaxies, then it will also affect the emissivity.  In the
present work we assume that $\mu={\rm const}=0.61$ at all radii, but one
should bear in mind that the results might change if the helium abundance
changes with radius or simply differs from the canonical value of
$7.92\times10^{-2}$ (number density relative to hydrogen) adopted here.

\subsection{Deviations from spherical symmetry}
\label{sec:ns}
To make a crude estimate of an impact on $v_c$ by possible
deviations from spherical symmetry one can make deprojection
analysis in individual wedges and compare the results. We divided the
data on each object into two independent halves - Northern and
Southern and repeated the analysis for each half separately. The 
difference in the circular speeds between two halves is given in
Table \ref{tab:errx} (column $\Delta_{\rm NS}$):
\begin{eqnarray} 
\Delta_{\rm NS}=\frac{v_{c,{N}}-v_{c,{S}}}{v_{c,{X}}},
\end{eqnarray} 
where $v_{c,N}$ and $v_{c,S}$ are the circular speeds for the Northern
and Southern halves respectively. For most of the objects the
difference is at the level of 2-4\% (compared to $\sim$1\%
pure statistical uncertainty). 

The largest difference $\sim$8\% is
for NGC1407. Inspection of the data have shown that this difference
primarily comes from XMM-Newton data. We note that NGC1407 was observed by
XMM-Newton with an offset angle of $\sim 8'$ and the PSF distortions
might contribute to this difference.

In the analysis of optical data (\S\ref{sec:robust}) we discuss
optical constraints on the gravitational potential, assuming that the
galaxies are spherically symmetric. All objects in the sample are
round, being E0-E2 galaxies, although in some objects (e.g. M87) the
ellipticity increases with radius. We note here that in the
radial range where stars are dominating the mass, the potential is
more spherical than the distribution of the stars themselves. The same
is likely true even in outer regions, given a steep decrease
($R^{-2}$ or steeper) of the optical surface brightness.

The analysis of independent wedges is not a useful indicator of a
possible error in the effective circular speed when the object posses
an axial symmetry and is viewed \mbox{pole-on}. Possible effects of a
non-spherical potential on optical estimates for a \mbox{pole-on} galaxy are
discussed in C08 (Section 7.2 there). There we show that for an axis
ratios (along and perpendicular to the line-of-sight) of 0.5 (oblate)
and 2 (prolate) the error in mass is factor of 0.79 and 1.04
respectively\footnote{See C08 for the assumptions used in this
calculation}, and argue that the probability that many objects in our
sample are strongly oblate or oblate systems viewed pole-on is rather
low. Therefore on average the error is going to be smaller than these
extreme values.

X-ray analysis of the ellipsoidal objects is discussed in
\citealt{2003A&A...398...41P} and C08. If potential is logarithmic and
the isothermal gas is in hydrostatic equilibrium, then the potential
is recovered correctly from the spherically symmetric analysis (see
Appendix B in C08). For a more complicated potential or non-isothermal
gas the deviations are present \citep[see e.g. Fig.1 in ][for an
example of expected bias for prolate and oblate cases in A2390
cluster]{2003A&A...398...41P}. Although the effects of non-spherical
potential are different for X-ray and optical analysis, the sign of
the effect for oblate and prolate systems is the same. Therefore
errors in mass/potential partly compensate each other when effective
circular speeds are compared.

We concluded that the assumption of a spherical potential can
contribute to the discrepancy between X-ray and optical data, but it
is unlikely that on average the magnitude of the effect exceeds
several per cent.

\subsection{Contribution of unresolved LMXBs and weak sources} 
\label{sec:lmxb}
The contribution of low-mass X-ray binaries (LMXB) can dominate the
X-ray emission in gas-poor elliptical galaxies
\citep[e.g.,][]{1985ApJ...296..447T}. Even when the spatial resolution and
sensitivity allows the resolution of some LMXBs the remaining emission can
still be dominated by even weaker unresolved sources, such as
accreting white dwarfs \citep[see, e.g.,][for the analysis of the 
  galaxy NGC3379]{2008A&A...490...37R}. 

Since all objects in our sample are gas-rich galaxies the
contribution of unresolved sources is not a major issue, especially
for such bright objects as M87 and NGC1399. For less luminous objects,
because of their harder spectrum compared to the hot gas emission, LMXBs
can affect the observed fluxes at the high end of the Chandra and XMM-Newton
energy bands. To test the magnitude of this effect we added a power law
component with a photon index $\Gamma=1.6$ with a free normalization
to represent the contribution of unresolved sources \citep[e.g.,][]{2003ApJ...587..356I,2006ApJ...653..207D,2006ApJ...639..136H}
and recalculated the circular speed for NGC4472. The resulting values are only 
$\sim$1\% lower than in Table \ref{tab:sigmax}. 

 Relative changes in the circular speed when a power law
component with a photon index $\Gamma=1.6$ is added to the spectral model
  for all objects in the sample are calculated in Table
  \ref{tab:errx} (column $\Delta_{\rm LMXB}$):
\begin{eqnarray} 
\Delta_{\rm LMXB}=\frac{v_{c,{\rm
      LMXB}}-v_{c,{X}}}{v_{c,{X}}}.
\end{eqnarray} 
On average adding a power law component shifts the circular speed few
percent lower. The largest effect is for M87 and in this very gas
rich galaxy the change in $v_c$ likely reflects the complexity of the
spectrum rather then the contribution of real LMXBs.

\subsection{Summary on uncertainties in X-ray analysis}
\label{sec:errx}
The summary of the uncertainties in determination of the
circular speed from X-ray data is given in Table
\ref{tab:errx}. The columns in the table show the relative deviation
of the circular speed from the reference value $v_{c,{X}}$ given
in Table \ref{tab:sigmax} when changes are made to the analysis
procedure. 
  The quoted uncertainties in columns labeled \"Err.\" are
pure statistical errors. They have clear meaning only when the
difference between the Northern and Southern parts of the galaxies are
considered, since the data are independent. In all other cases the quoted
uncertainty corresponds to the largest statistical error in one
of the two values of $v_c$ used to calculate $\Delta$.

The last two rows in the Table \ref{tab:errx} give the mean change in
the circular speed and the RMS value (relative to zero). The values
given in the Table allows one to get an idea of the uncertainties
introduced by the assumptions incorporated into calculations of
$v_c$. For instance, letting abundance be a free parameter on average shifts
the value of $v_c$ up by $\sim 2$\%, while adding a power law
component (to control possible contribution of LMXBs) on the contrary
shifts $v_c$ down by $\sim 2$\%. Changing any of the bounds of the
fitting range $r_1$ and $r_2$ by a factor of 2 changes $v_c$ by few \%
up or down.

While rigorous evaluation of combined uncertainties is difficult, we
can very crudely estimate it by adding RMS values of $\Delta_{\rm
  abund}$, $\Delta_{\rm LMXB}$, $\Delta_{\rm NS}$, $\Delta_{\rm r1}$
and $\Delta_{\rm r2}$ quadratically. The resulting value 7.1\% is
crude a characteristic of the uncertainties in the final value of
$v_{c,X}$ introduced by the modifications of our analysis procedure. Note that
 pure statistical errors also contribute to the above estimate.

\begin{table*}
\centering
\caption{\label{tab:errx} Relative changes in circular speed with
  respect to the reference value $v_{c,{X}}$ given
in Table \ref{tab:sigmax} when changes are made to the analysis
procedure (see \S\ref{sec:errx}).}
\begin{tabular}{rrrrrrrrrrr}
\hline
Galaxy &  $\Delta_{\rm abund}$ & Err. &  $\Delta_{\rm LMXB}$ & Err. &
$\Delta_{\rm NS}$ & Err. &  $\Delta_{\rm r_1\times 2}$ & Err.&
$\Delta_{\rm r_2/2}$ & Err.\\
& \% &  & \% & &\% & & \% & & \% & \\
\hline
ngc1399 &  4.23 &  0.80 &  -2.01 &  0.44 &  2.36 &  0.64 &  1.40 &  0.94 &  1.22 &  0.39\\
ngc1407 &  0.76 &  5.67 &  -1.61 &  2.37 &  8.25 &  2.88 &  5.66 &  1.74 &  1.05 &  2.22\\
ngc4472 &  2.43 &  2.17 &  -0.98 &  0.79 & -2.31 &  1.16 & -0.42 &  0.76 & -3.71 &  0.70\\
ngc4486 &  3.88 &  0.40 &  -5.69 &  0.37 & -4.58 &  0.39 &  1.44 &  0.26 & -1.00 &  0.31\\
ngc4649 & -2.48 &  1.14 &   0.36 &  0.41 &  2.67 &  0.62 &  3.30 &  0.43 &  0.08 &  0.39\\
ngc5846 &  1.53 &  2.61 &  -1.82 &  1.02 & -2.57 &  1.50 & -1.68 &  0.81 & -5.14 &  1.21\\
\hline
Mean &  1.73 && -1.96 &&  0.64 &&  1.62 && -1.25 \\
RMS &  2.82 &&  2.69 &&  4.35 &&  2.89 &&  2.70 \\
\hline
\end{tabular}
\begin{flushleft}
\hspace{1.92cm}
\begin{tabular}{lll}
\hline
Name & changes in the analysis & Section \\
\hline
$\Delta_{\rm abund}$ & free metal abundance & (\S\ref{sec:metals}) \\
$\Delta_{\rm LMXB}$ & a power law is added &(\S\ref{sec:lmxb})\\
$\Delta_{\rm NS}$ & difference between North and South &(\S\ref{sec:ns})\\
$\Delta_{\rm r1}$ & $r_1\times 2$ & (\S\ref{sec:sample})\\
$\Delta_{\rm r2}$ & $r_2/2$ & (\S\ref{sec:sample})\\
\hline
\end{tabular}
\end{flushleft}
\end{table*}

\section{Robust estimators of the isothermal potential from optical data} 

\label{sec:robust}

The results of the X-ray analysis suggest that the shape of the potential
profile of the galaxies in our sample is not far from isothermal
over the range of radii $\sim 0.5$--25 kpc, corresponding to $\sim 0.05$--$3~R_e$. Let us assume that
this is true at all radii and the total gravitational potential of a
galaxy is logarithmic, 
\begin{eqnarray}
\varphi(r)=v_c^2 \log r + {\rm const.}
\end{eqnarray}
It is natural to ask, given this assumption, how the velocity dispersion of
the stars in the galaxy is expected to be related to $v_c$. A robust method
to determine $v_c$ from optical observations would enable us to check our
results by comparing the derived values in Table \ref{tab:sigmax}, to
determine $v_c$ and thus the gravitational potential for many galaxies without
X-ray observations, and to look for evidence of non-thermal pressure support
in the X-ray gas.  

For simplicity we shall assume that the galaxy is spherical with known surface
brightness $I(R)$ and line-of-sight velocity dispersion $\sigma(R)$, and ask 
what is the best way to determine $v_c$ from known $I(R)$, $\sigma(R)$ given
that the potential is isothermal. 

\subsection{Circular speed from velocity dispersion for distant (unresolved) galaxies}

Since the acceleration of a star in the logarithmic potential $\ddot{\mathbf
  r}= -\bnabla \varphi=v_c^2 {\mathbf r}/r^2$, the virial theorem for
such a system reads
\begin{eqnarray}
0=\left < \dot{{\mathbf r}}^2+{\mathbf r}\cdot\ddot{\mathbf
  r}\right > = \left <  {\mathbf
  v}^2\right >-v_c^2, 
\end{eqnarray}
where ${\mathbf v}=\dot{{\mathbf r}}$. Since the galaxy is spherical,
the mean-square velocity along any line of sight is equal to
$\frac{1}{3}\langle {\mathbf v}^2\rangle$ and therefore 
\begin{eqnarray}
v_c^2=3\,\frac{\int_0^\infty dR R\, I(R)\sigma^2(R)}{\int_0^\infty dR\, R
  I(R)},
\label{eq:virial}
\end{eqnarray}
i.e., the circular speed is equal to $\surd{3}$ times the integrated 
line-of-sight dispersion for the whole galaxy. 
This formula is useful for distant (unresolved) galaxies that lie
entirely within the spectrograph slit, or  within the
field of an integral-field spectrograph. For nearby galaxies, however,
this formula is not useful since velocity-dispersion profiles
typically go out only to the effective radius $R_e$ (eq.\ \ref{eq:reff}),
which contains only half the light. 

\subsection{Circular speed from velocity dispersion for S\'ersic models}

\subsubsection{Constant anisotropy}
\label{sec:sweet}

\begin{figure*}
\plotthree{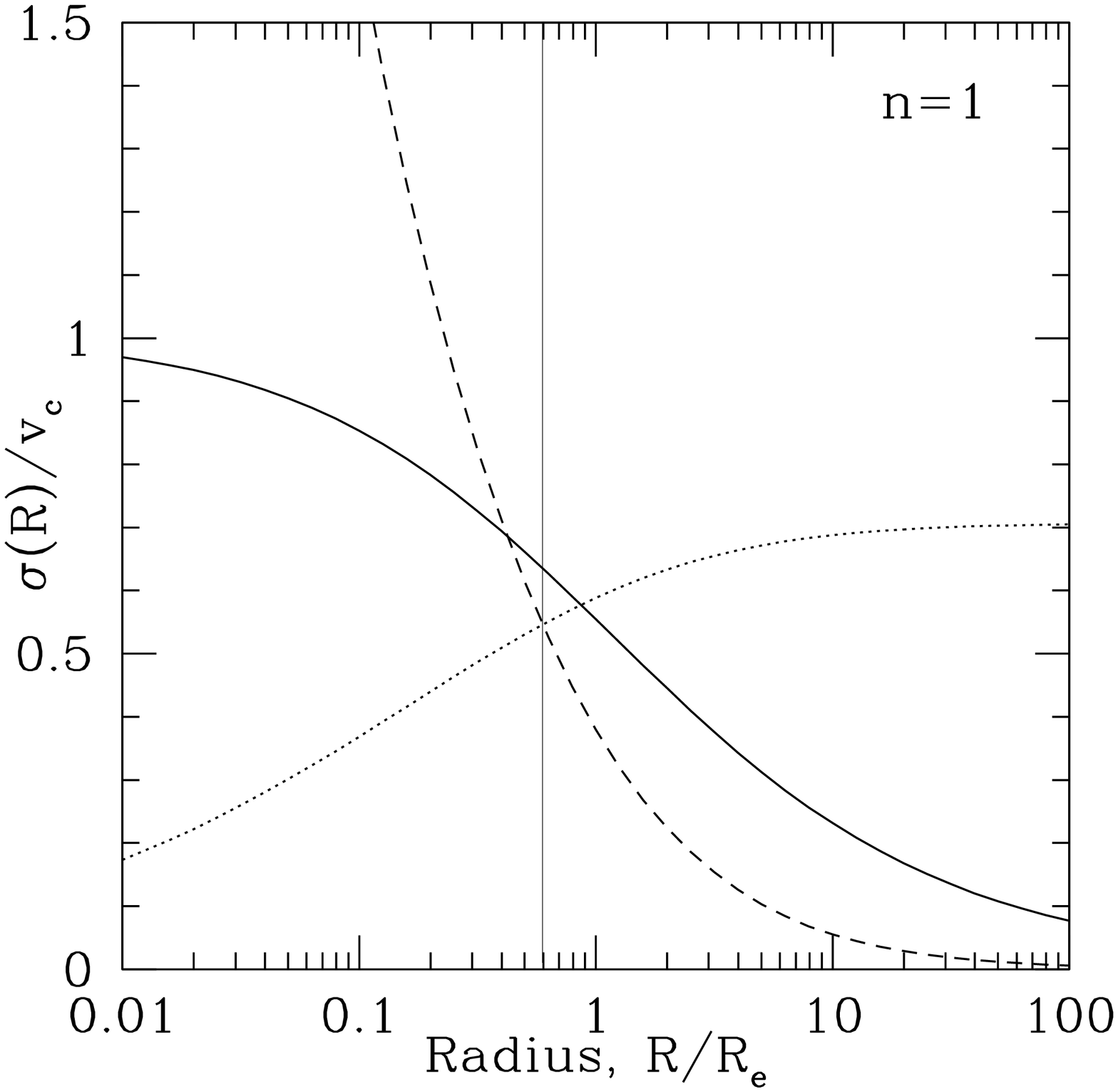}{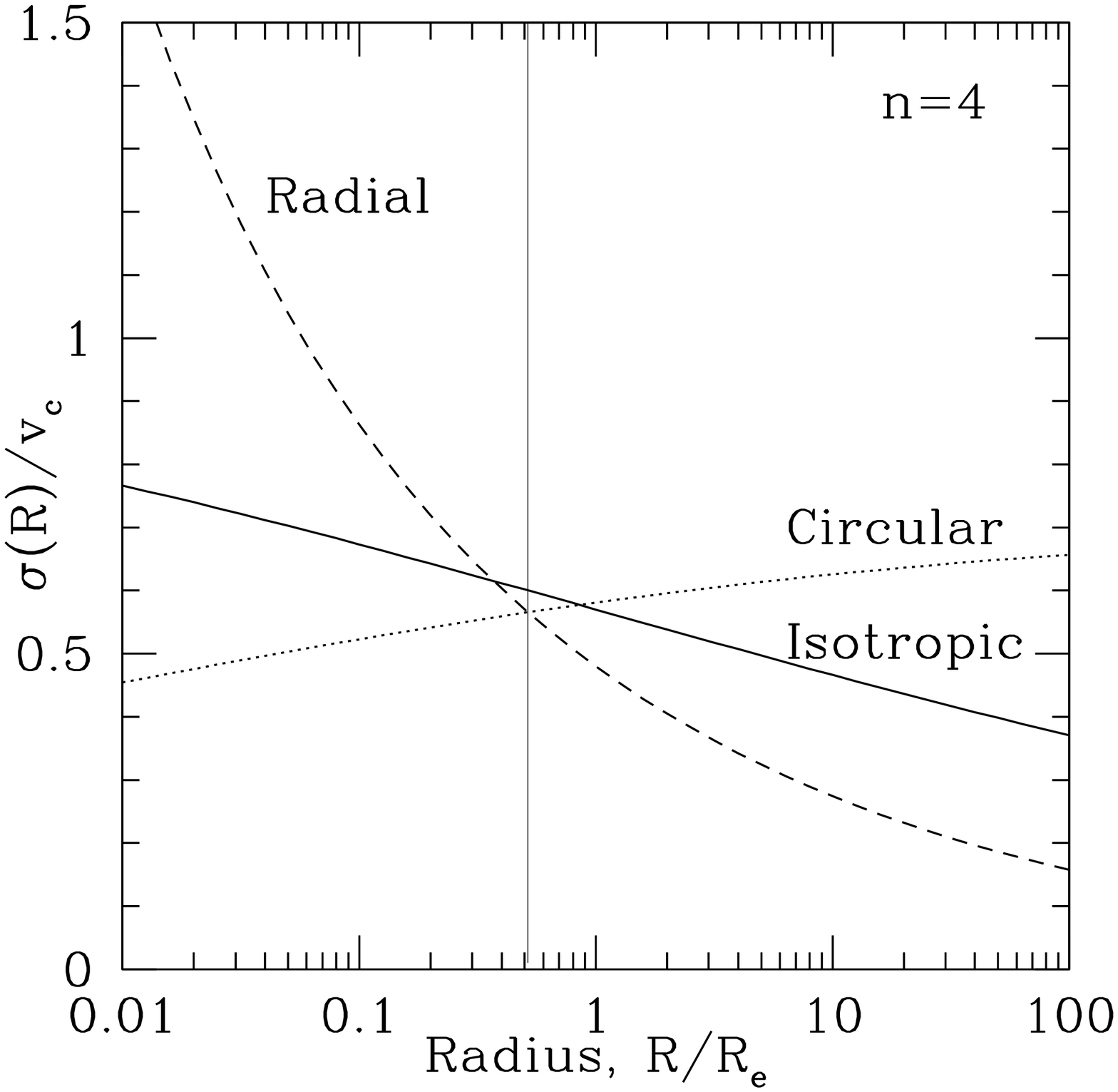}{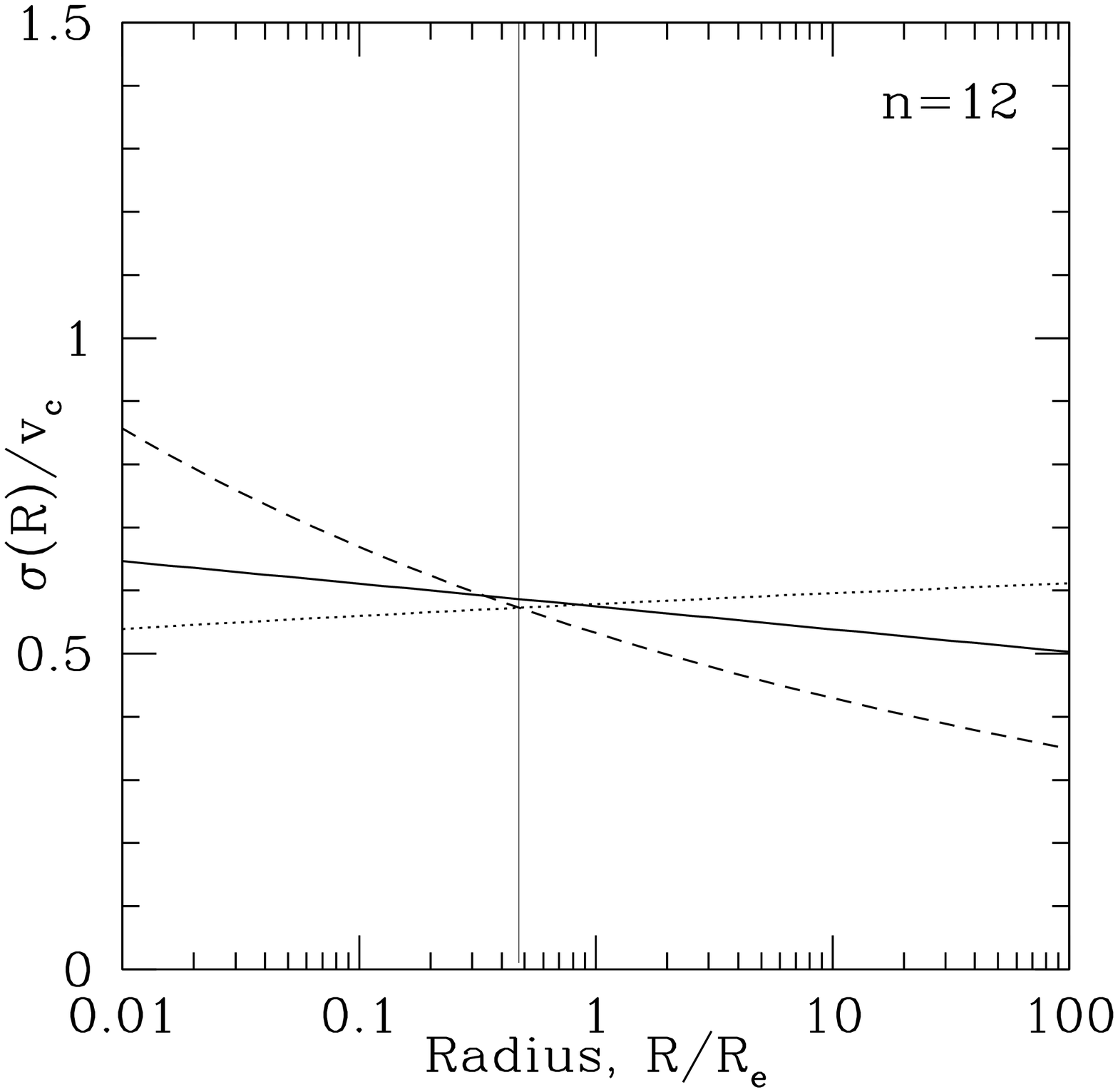}
\caption{Line-of-sight velocity dispersion as a function of radius for
spherical galaxy models with S\'{e}rsic index $n=1,4,12$ -- left, middle,
right panels respectively. The solid, dotted and dashed lines correspond to
systems composed of isotropic, circular and radial stellar orbits.  The thin
vertical line shows the position of the ``optimal radius'' or ``sweet spot'' at which the 
deviations among the three models are minimized. 
\label{fig:s2n}
}
\end{figure*}

Let us continue to assume that we have a spherical galaxy in a logarithmic
potential and consider three variants of the stellar orbital anisotropy:
isotropic orbits (i.e., the velocity-dispersion tensor is isotropic), 
circular orbits, and radial orbits. It is straightforward to show from the
Jeans equation (see Appendix \ref{ap:jeans}) that the line-of-sight velocity
dispersion in these three models is given by:
\be
\sigma^2(R)=\sigma_{\rm iso}^2(R) \equiv v^2_c\frac{R}{I(R)}\int_R^\infty
\frac{I(x)}{x^2}dx,
\label{eq:siso}
\ee
for isotropic orbits,
\be
\sigma^2(R)=\sigma_{\rm circ}^2(R) \equiv \frac{1}{2}\left [v^2_c
  -\sigma_{\rm iso}^2(R) \right ],
\label{eq:scirc}
\ee
for circular orbits and
\be
\sigma^2(R)=\sigma_{\rm rad}^2(R) \equiv \frac{1}{2}
v^2_c\frac{1}{RI(R)}\int_R^\infty I(x)\left(1 - R^2/x^2 \right)dx,
\label{eq:srad}
\ee
for pure radial orbits\footnote{Here we ignore the possibility that some of
  the  models considered may have unphysical (negative) distribution functions
  at some radii. This is, for example, the case for a pure radial-orbit model
  when $r\rightarrow 0$ so long as the emissivity diverges more slowly than
  $j(r)\propto r^{-2}$ \citep{1984ApJ...286...27R}.}.  In Figure \ref{fig:s2n}
the radial dependence of the line-of-sight velocity dispersion is shown for
the above three cases, assuming that the surface brightness of the galaxy is
described by a S\'{e}rsic law $I(R)\propto \exp(-a R^{1/n})$ with index
$n=1,4,12$. The radius is plotted in units of the effective radius $R_e$, the
radius containing half the light:
\be
\int_0^{R_e} RI(R)dR=\frac{1}{2} \int_0^{\infty} RI(R)dR.
\label{eq:reff}
\ee
{}From Figure \ref{fig:s2n} it is clear that for all $n$ and for all three
models spanning the range of possible anisotropies the dispersions are rather
similar at about 0.5--$0.6 R_e$. This result suggests that somewhere near this
radius the line-of-sight velocity dispersion may provide a measure of the
circular speed of the underlying isothermal potential that is relatively
independent of the details of the stellar velocity distribution. 

\begin{figure}
\plotone{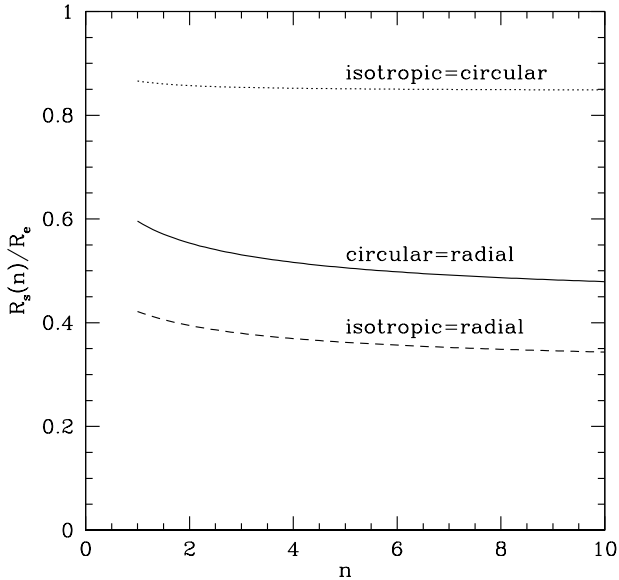}
\caption{Radius $R_s(n)$ at which the line-of-sight dispersions
  (eqs.\ \ref{eq:siso}, \ref{eq:scirc}, and \ref{eq:srad}) for
  systems composed from different types of orbits coincide, as a function
  of S\'ersic index $n$. Solid
  line: $\sigma_{\rm circ}(R)=\sigma_{\rm rad}(R)$, dashed line:
  $\sigma_{\rm iso}(R)=\sigma_{\rm rad}(R)$, dotted line:
  $\sigma_{\rm iso}(R)=\sigma_{\rm circ}(R)$. 
\label{fig:sweet}
}
\end{figure}

\begin{figure}
\plotone{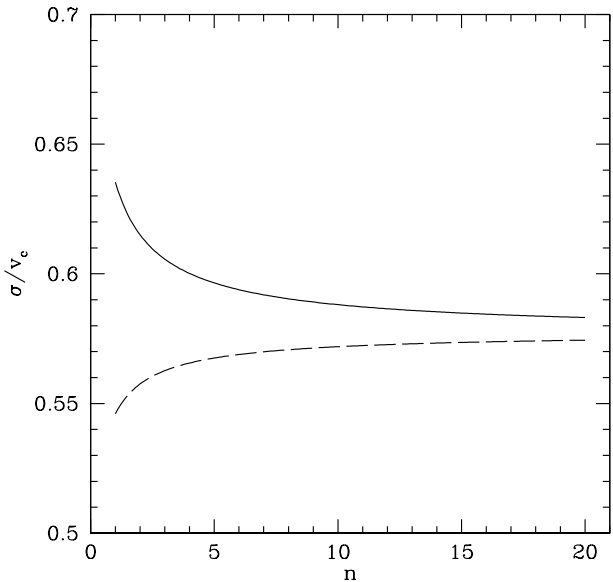}
\caption{The line-of-sight
  dispersions evaluated at $R_s(n)$ [where
  $\sigma_{\rm circ}(R_s)=\sigma_{\rm rad}(R_s)$] are shown with solid
  (for systems with intrinsically isotropic orbits)
  and dashed (radial and circular orbits) lines. 
\label{fig:svc}
}
\end{figure}

There are many possible ways to define the ``optimal'' radius or ``sweet
spot'' at which the relation between the line-of-sight velocity dispersion and
the circular speed is likely to exhibit the smallest variation. In Figure
\ref{fig:sweet} we plot the radii where the line-of-sight dispersions from
each pair of the three models coincide, as a function of S\'ersic index $n$.
{}From equations (\ref{eq:siso}), (\ref{eq:scirc}) and (\ref{eq:srad}) these are
the solutions of the equations
\be
\begin{array}{ll}
n\Gamma(n,z)=z^ne^{-z}&{\rm circular=radial} \\
3 z^{n}n \Gamma(-n,z)=e^{-z}&{\rm isotropic=circular} \\
3 z^{2n}\Gamma(-n,z)=\Gamma(n,z)&{\rm isotropic=radial},
\end{array}
\label{eq:rsweet}
\ee
where $z=aR_s^{1/n}$, and $\Gamma$ is the upper incomplete gamma function.
Models of galaxy formation usually suggest that the velocity-dispersion
tensor is isotropic near the center and radially biased
($\sigma_r^2>\sigma_\theta^2$) in the outer parts of the galaxy, so we
should prefer the bottom half of the range of $R_s$ indicated by these curves.
A natural choice for the sweet spot $R_s$ is the middle curve, where the
line-of-sight velocity dispersions for radial and circular models are equal
(thin vertical lines in Figure \ref{fig:s2n}). This can be written for
arbitrary surface-brightness profiles as
\begin{eqnarray}
\int_{R_s}^{\infty}I(x)dx=R_sI(R_s).
\label{eq:sweetcr}
\end{eqnarray}
The values of $R_s$, $\sigma_{\rm iso}(R_s)$, $\sigma_{\rm circ}(R_s)$ and
$\sigma_{\rm rad}(R_s)$ (the latter two quantities coincide by the definition
of $R_s$) are shown in Figure \ref{fig:svc} and in Table \ref{tab:rs}, as a
function of $R_s/R_e$.

The range of dispersions spanned by these three models is of course smaller
than the range spanned by all possible equilibrium models (see for example
Figures 1 and 2 of \citealt{1984ApJ...286...27R}). Our results nevertheless
suggest that the line-of-sight velocity dispersion evaluated at $R_s\simeq
0.5R_e$ is a better proxy for the circular speed in a galaxy with a
logarithmic potential than the dispersion at any other radius (e.g., the
central velocity dispersion).

Determination of the effective radius and the S\'{e}rsic index
of real galaxies depends (sometimes strongly) on the range of radii
used to fit the surface brightness profile and on the method of extrapolating
to calculate the total flux. This is especially true
for galaxies with extended envelopes, characteristic of central
galaxies in clusters and groups. For such galaxies the reported values of
$R_e$ and $n$ for the same object may vary strongly. A striking
example is M87: the effective radius of $145''$ in Table \ref{tab:sample} 
\citep{1994MNRAS.271..523D} is a factor of almost five smaller than the
most recent value $R_e=704''$ in
\cite{2009ApJS..182..216K}. Similarly the reported values of the
S\'{e}rsic index vary from 6.51 in
\cite{1994MNRAS.271..523D} to 11.84 in \cite{2009ApJS..182..216K}. The
position of the sweet spot changes accordingly. Clearly such large
variations have two main causes: (i) deviations of the observed
surface-brightness profile from a pure S\'{e}rsic profile, and (ii) the weak
dependence of the surface-brightness slope on $R/R_e$ and $n$ or, more
precisely, strong covariance between $R_e$ and $n$ when fitting to data. Indeed
for the S\'{e}rsic profile the slope is \citep[e.g.,][]{2005PASA...22..118G}
\begin{eqnarray} 
-\frac{d\log I(R)}{d\log R}\approx 2 \left(\frac{R}{R_e}\right)^{1/n}.  
\label{eq:replacethis} 
\end{eqnarray}
Obviously when $n$ is large the variations of slope at a given $R$
are not very sensitive to the variations of $R_e$.

This raises the question of which set of $n$ and $R_e$ to use and how
strongly the final result (circular speed evaluated near the sweet
spot) depends on these parameters. The definition of the sweet spot
(eq.\ \ref{eq:sweetcr}) does not explicitly use the value of $R_e$, but
is instead most sensitive to the slope of the surface brightness
profile; thus, for example, if
$I(R)\propto R^{-2}$ for $R>R_0$ then any radius $R>R_0$ can equally well
serve as the sweet point (see also \S\ref{sec:local}).  This implies
that the evaluated circular speed should not be sensitive to the precise
values of $R_e$ and $n$ so long as they provide a good fit to the
surface-brightness profile and our other assumptions (e.g., circular speed
independent of radius) are satisfied. 

To illustrate this point, let us assume that M87 has circular speed
independent of radius, isotropic velocity-dispersion tensor, and
surface-brightness profile that is accurately fit by a S\'ersic profile with
the Kormendy et al.\ parameters, $R_e=704''$ and $n=11.84$, and that we estimate
the circular speed using Table \ref{tab:rs} and the D'Onofrio et
al. parameters $R_e=145"$ and $n=6.51$. The resulting error in $v_c$ will be
only 3\% even though the values of $R_e$ differ by a factor of almost five. 

In practice the sweet spot at large
radius is less favorable since measurements of the line-of-sight
velocity dispersion have only a limited radial extent. In particular, to
estimate the circular speed using the effective radius from
Kormendy et al.\ one needs the stellar velocity dispersion
at a radius of $\sim 0.5 R_e \sim5$--$6'$, which is not known. For this
reason we decided to use the effective $R_e=145''$ of D'Onofrio et al.

It is possible that equation \ref{eq:sweetcr} does not have roots at
radii where good measurements of the line-of-sight velocity dispersion
are available. In this case one can look for the radius $R$ with
measured $\sigma(R)$ where the relative mismatch of the left and right sides of
equation \ref{eq:sweetcr} is smallest. 
 
In the limit of large S\'ersic index $n$, the surface brightness of a galaxy at
radii not too far from the effective radius declines as $R^{-2}$
\citep[e.g.,][]{2005PASA...22..118G} and therefore volume emissivity declines
as $r^{-3}$. While formally the total mass and the effective radius diverge as
$n$ goes to infinity, at the same time the line-of-sight velocity dispersion
becomes independent of the radius and the anisotropy of stellar orbits
\citep{1993MNRAS.265..213G} as long as the shape of the velocity-dispersion
tensor, $\sigma_\theta^2/\sigma_r^2$,  does not depend on radius. In this limit
$\sigma^2(R)={\rm const}=v^2_c/3$ at all radii (Table \ref{tab:rs}).

\begin{table}
\centering
\caption{\label{tab:rs}Dependence of $R_s$ and line-of-sight velocity
  dispersions at $R_s$ on the S\'ersic index $n$. Here $R_s$ is such
  that $\sigma_{\rm circ}(R)=\sigma_{\rm rad}(R)$; $R_s$ is measured in units
  of the effective radius $R_e$, and $\sigma$ is measured in units of the
  circular speed $v_c$. }
\begin{tabular}{llllll}
\hline
$n$   &$R_s/R_e$ & $\sigma_{\rm iso}$& $\sigma_{\rm circ}=
\sigma_{\rm rad}$ & $(\sigma_{\rm iso}-\sigma_{\rm circ})/\sigma_{\rm iso}$\\
\hline
1    &   0.595824   &   0.635337   &   0.546053   &   0.140530    \\
2    &   0.553551   &   0.615010   &   0.557567   &   0.0934008   \\
3    &   0.530947   &   0.605640   &   0.562672   &   0.0709465   \\
4    &   0.516328   &   0.600140   &   0.565611   &   0.0575356   \\
5    &   0.505918   &   0.596490   &   0.567538   &   0.0485364   \\
6    &   0.498051   &   0.593877   &   0.568907   &   0.0420463   \\
7    &   0.491859   &   0.591909   &   0.569932   &   0.0371287   \\
8    &   0.486838   &   0.590369   &   0.570730   &   0.0332661   \\
9    &   0.482674   &   0.589130   &   0.571369   &   0.0301473   \\
10   &   0.479156   &   0.588111   &   0.571894   &   0.0275738   \\
11   &   0.476140   &   0.587256   &   0.572333   &   0.0254124   \\
12   &   0.473523   &   0.586530   &   0.572705   &   0.0235703   \\
13   &   0.471228   &   0.585904   &   0.573025   &   0.0219810   \\
14   &   0.469199   &   0.585359   &   0.573304   &   0.0205953   \\
15   &   0.467389   &   0.584881   &   0.573548   &   0.0193760   \\
16   &   0.465765   &   0.584457   &   0.573764   &   0.0182946   \\
17   &   0.464299   &   0.584078   &   0.573957   &   0.0173289   \\
18   &   0.462968   &   0.583738   &   0.574130   &   0.0164609   \\
19   &   0.461754   &   0.583432   &   0.574285   &   0.0156766   \\
20   &   0.460642   &   0.583153   &   0.574427   &   0.0149643   \\
$\infty$&$\infty$   &   0.577350   &   0.577350   &   0           \\
\hline
\end{tabular}
\end{table}

\subsubsection{Anisotropy changing with radius}
In the models considered above the anisotropy parameter was independent of 
radius. We now consider several simple models in which anisotropy
parameter $\beta(r)=1-\sigma_\theta^2(r)/\sigma_r^2(r)$ depends on
radius according to 
\begin{eqnarray}
\beta(r)=\frac{\beta_2r^c+\beta_1r_a^c}{r^c+r_a^c~},
\label{eq:om}
\end{eqnarray}
where $\beta_1$ and $\beta_2$ are the anisotropy parameters at $r=0$ and
$r\to\infty$ respectively, $r_a$ is the anisotropy radius, and the exponent $c$
controls the sharpness of the transition. The radial velocity dispersion is then 
\begin{eqnarray}
\sigma_r^2(r)=\frac{v_c^2}{j(r)W(r)}\int_r^\infty \frac{j(x)W(x)}{x}dx,
\label{eq:om1}
\end{eqnarray}
where 
\begin{eqnarray}
W(x)=x^{2\beta_1}(x^c+r_a^c)^{2(\beta_2-\beta_1)/c}.
\label{eq:om2}
\end{eqnarray}
A similar model with $c=1$, $\beta_1=0.2$, $\beta_2=1$ was used to describe
the anisotropy profile in M87 by \cite{doherty09}. Another special case of our
parametrization is the Osipkov--Merritt model
\citep{1979SvAL....5...42O,1985AJ.....90.1027M}, corresponding to
$c=2$, $\beta_1=0$, $\beta_2=1$.

\begin{figure*}
\plottwo{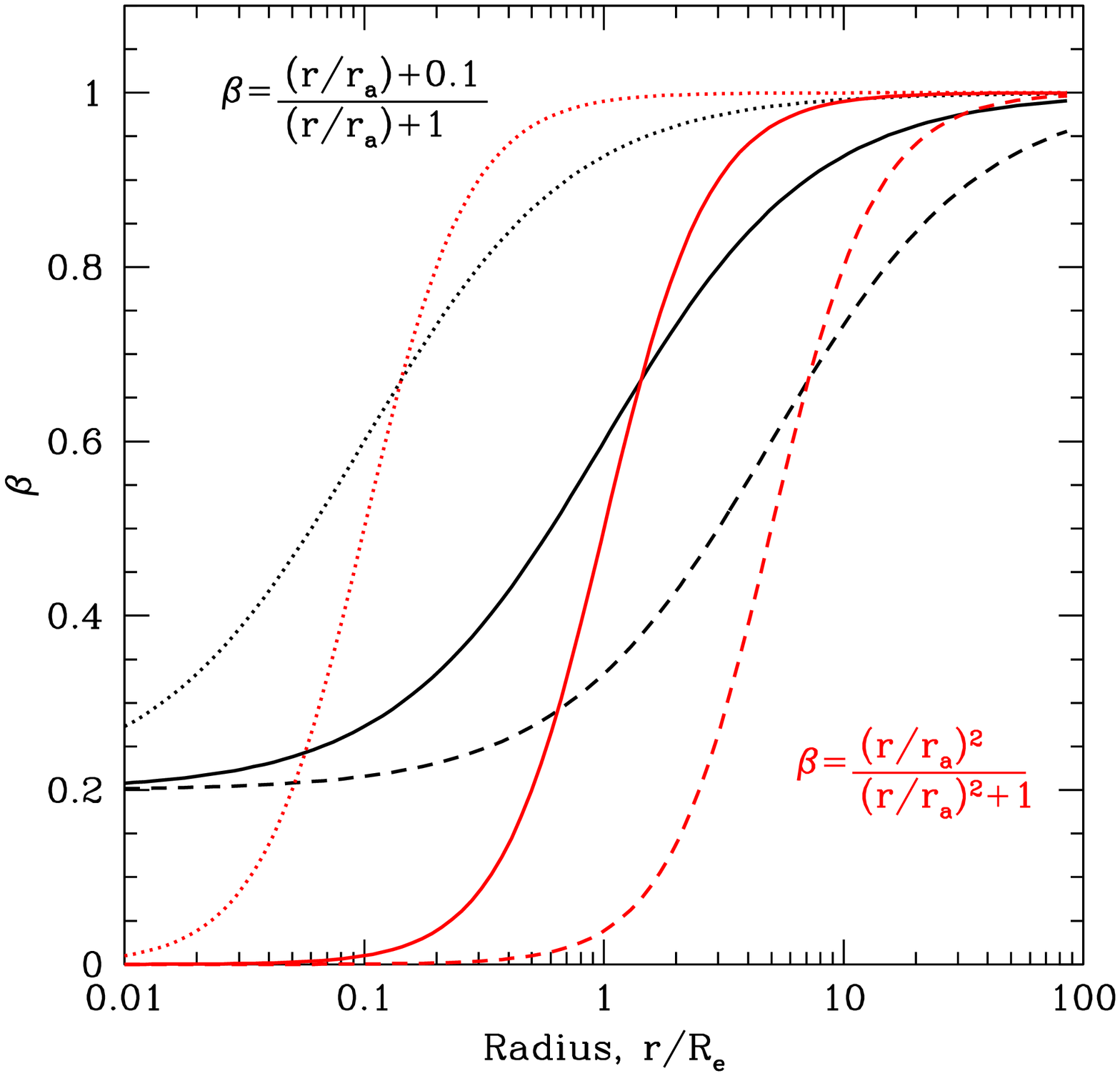}{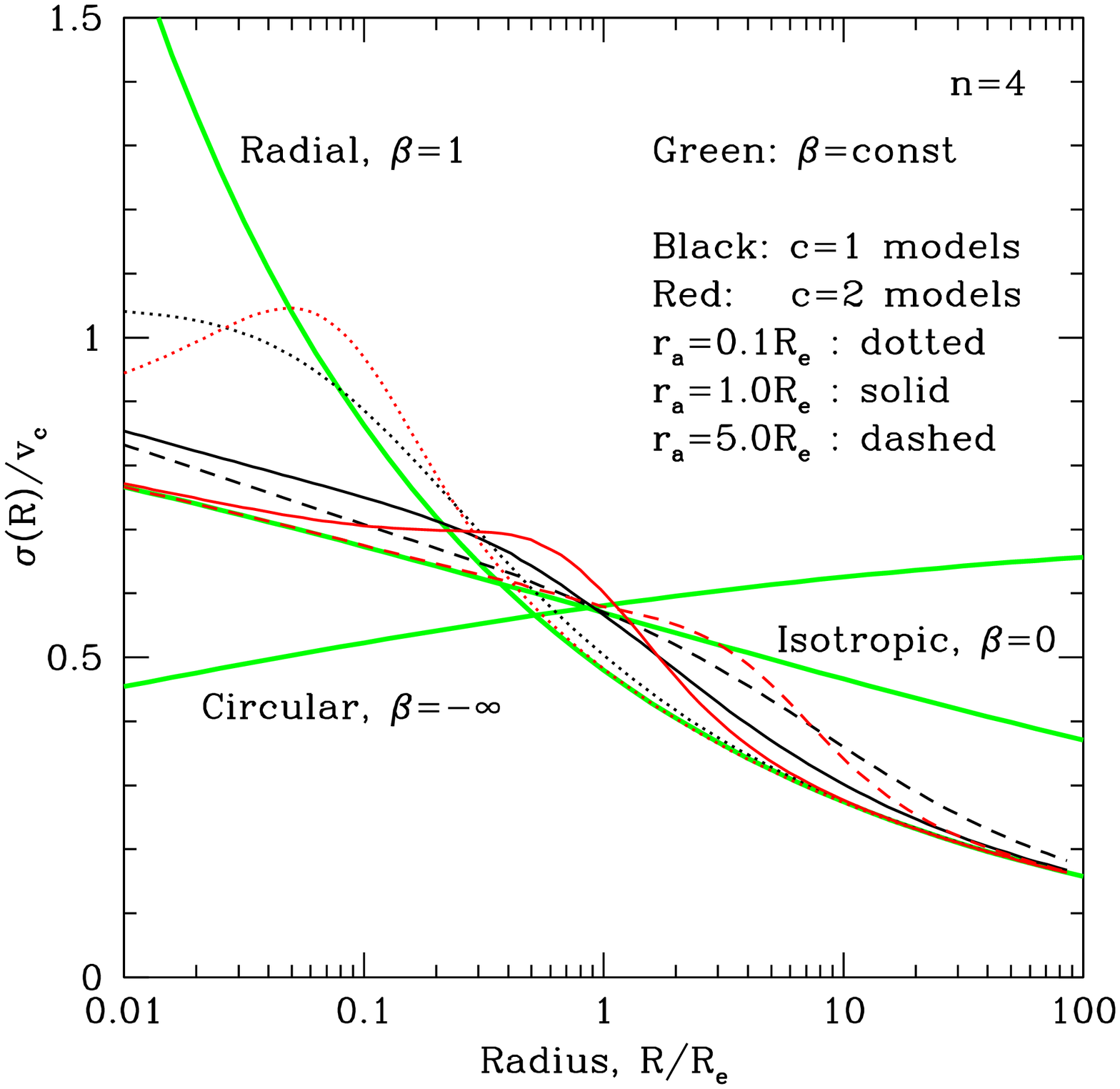}
\caption{Left: Radial dependence of the anisotropy parameter $\beta$.  The
Doherty et al.\ models ($c=1$) are shown by black lines, and the
Osipkov-Merritt models ($c=2$) by red lines. The curves
corresponding to different values of the anisotropy parameter $r_a$ (see
eq.\ \ref{eq:om}) are shown with  dotted, solid and dashed lines for
$r_a=0.1 R_e$, $1 R_e$ and $5 R_e$ respectively.
Right: Corresponding profiles of line-of-sight velocity
dispersion for an $n=4$ S\'ersic
model. The legends of the lines are the same
as in the left panel. The thick green lines show for reference our standard
isotropic, circular and radial models. 
\label{fig:om}
}
\end{figure*}

In the left panel of Figure \ref{fig:om} we show the behavior of the
anisotropy parameter $\beta$ for $c=1,2$ and 4 and
different values of the anisotropy radius $r_a$. The Doherty et al.\ models
($c=1$) are shown by black lines and  the Osipkov-Merritt
models ($c=2$) by red lines, both for $r_a=0.1 R_e$, $1 R_e$, $5 R_e$.
In the right panel of Figure
\ref{fig:om} we show corresponding profiles of line-of-sight velocity
dispersion for an $n=4$ S\'ersic model. The legends of the lines are the same
as in the left panel. For reference, the thick green lines show our standard isotropic,
circular and radial models.

We did not check explicitly that these solutions of the Jeans equation
correspond to non-negative distribution functions. S\'ersic models with
$\beta=1$, corresponding to purely radial orbits, certainly have negative
distribution functions and thus are unphysical. For the Osipkov--Merritt
models, the smallest anisotropy radius shown in Figure \ref{fig:om} is
$r_a=0.1R_e$, close to but outside the 
range $r_a\lesssim 0.05R_e$ at which the distribution function is negative
\citep{1985AJ.....90.1027M}.

Considering the models shown in the right panel of Figure \ref{fig:om} it is
clear that the ``sweet spot'' at which the sensitivity of the line-of-sight
dispersion to the orbit structure is minimized is near $0.5R_e$, just as we
found in \S\ref{sec:sweet} from examining models with radial, circular, and
isotropic orbits. However, and not surprisingly, the models with radially
varying anisotropy span a larger range in line-of-sight dispersion at the
sweet spot, with the largest deviations occurring when $r_a/R_e\sim 1$.  For
instance at $R=0.516R_e$ (taken from Table \ref{tab:rs} for $n=4$) the
Osipkov-Merritt model with $r_a=R_e$ predicts a line-of-sight velocity
dispersion 14\% larger than the isotropic model and  20\% larger than
a model composed of circular or radial orbits. Making the transition in $\beta$
smoother ($c=1$) or sharper ($c > 2$) makes the wiggle in the line-of-sight
velocity profile less (or more) pronounced as shown in Figure \ref{fig:om}.

So far we have discussed only systems with radial anisotropy (``Type I'' in the
notation of \citealt{1985AJ.....90.1027M}). We have not discussed ``Type II''
systems in which the orbits become predominantly circular at large radii,
because such systems are not produced in our current model of galaxy
formation. At least for the
Osipkov-Merritt models ($c=2$) these curves have an abrupt change in
the line-of-sight velocity dispersion at $R=r_a$ (see Figure 5b in
\citealt{1985AJ.....90.1027M}). These curves fill the area between the isotropic and
circular models and near $R=R_s(n)$ they predict smaller dispersion
than the isotropic model.

Even though spatially varying anisotropy can increase the spread of the
expected value of the line-of-sight velocity dispersion compared to our set of
isotropic, radial and circular models, the dispersion at the radius $R\simeq
0.5~R_e$ remains relatively insensitive to the orbital anisotropy and hence
its value at this radius provides a useful measure of the circular speed of
the underlying isothermal potential.

\subsubsection{Aperture dispersions and other extensions}

The methods we have described for estimating the circular speed $v_c$ rely on
the value of the line-of-sight velocity dispersion $\sigma(R)$ at a single
radius $R_s$. Obviously, one could imagine improved estimators based on some
combination of the dispersions at two or more radii $R_{s1}, R_{s2},\ldots$.
A related possibility is to use weighted averages of the form $\int_0^A
I(R)\sigma^2(R)f(R)\, dR/\int_0^A I(R)f(R)\, dR$ for some suitably chosen
function $f(R)$. Here we discuss only one possibility, the use of aperture
dispersions ($f(R)=R$), which we shall find to be {\em less} useful
than dispersions measured at a single radius.

\begin{figure*}
\plotthree{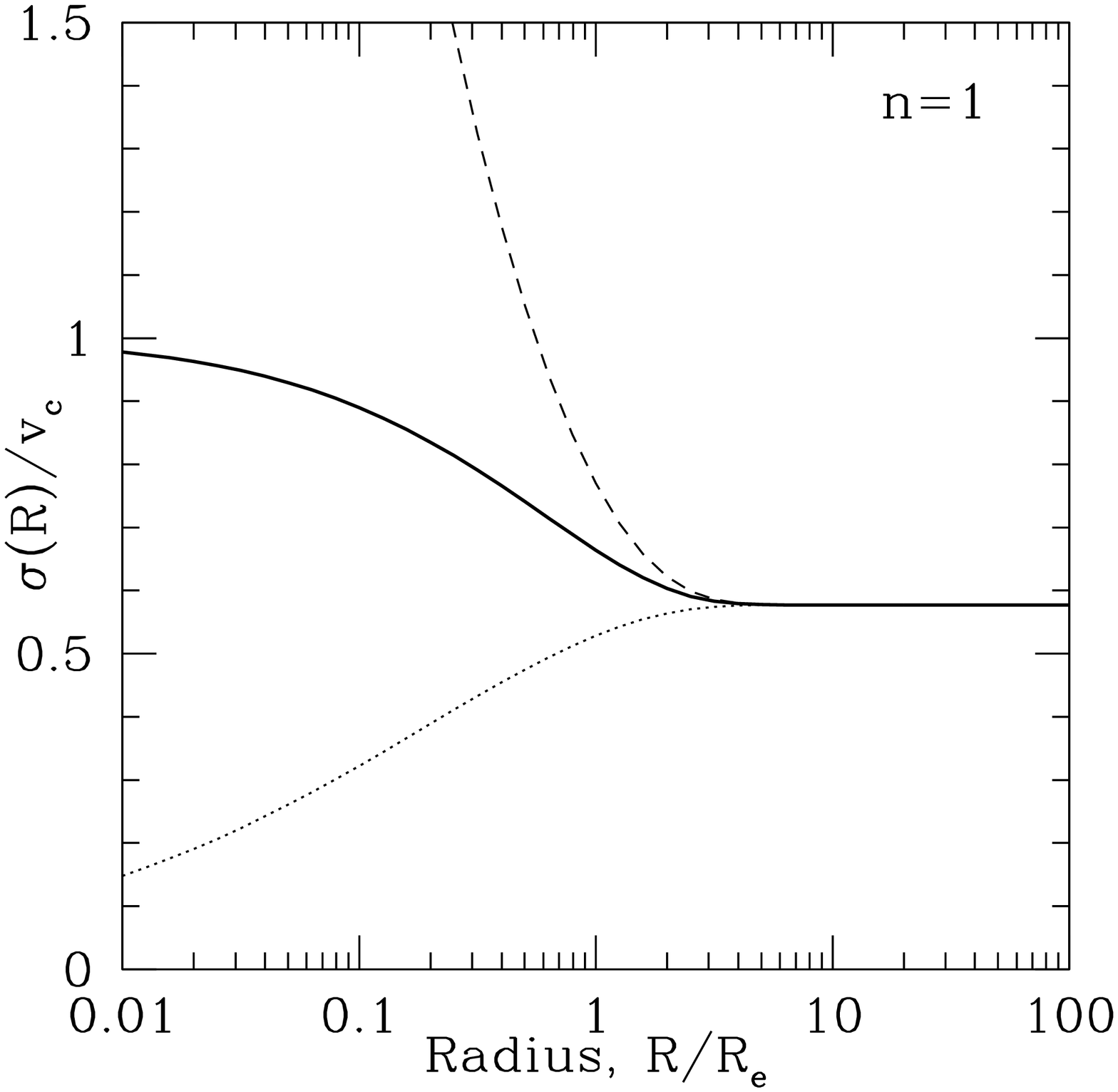}{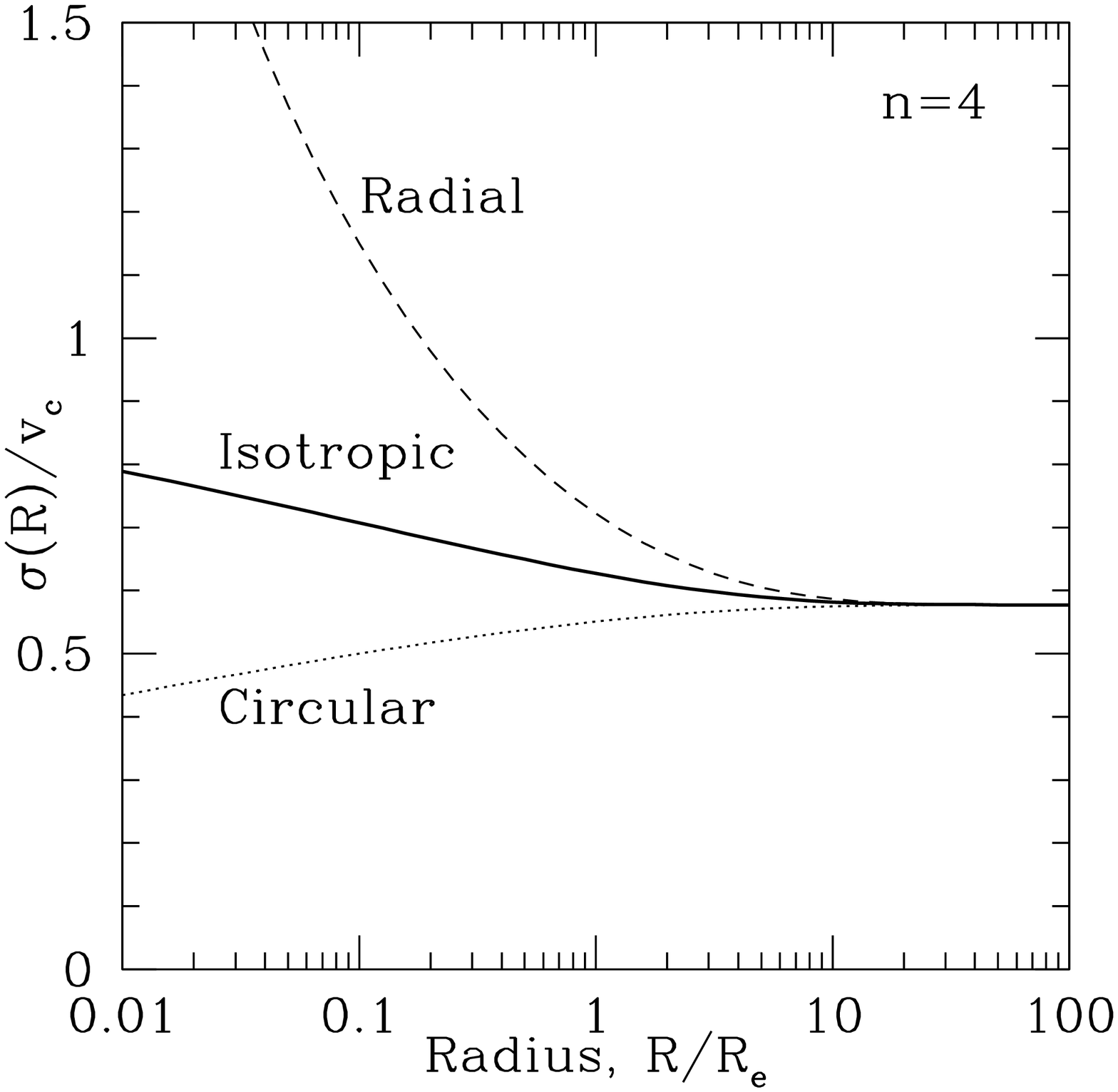}{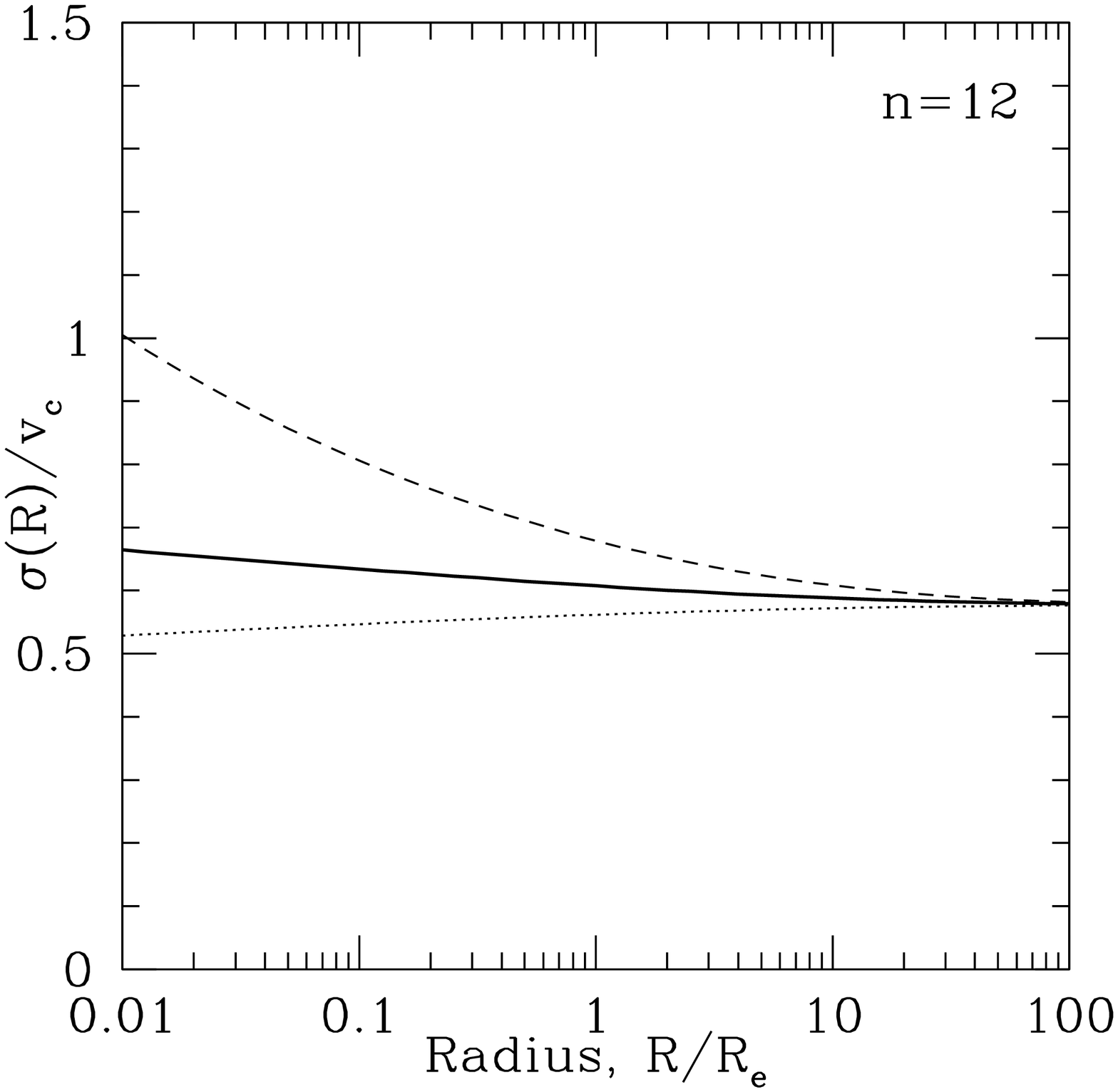}
\caption{Velocity dispersion within an aperture as a function of aperture radius for
  S\'{e}rsic index $n=1,4,12$. The solid, dotted, and  dashed lines correspond to
  systems composed from isotropic, circular and radial stellar orbits. 
\label{fig:aperture}
}
\end{figure*}

Similar to the derivations in Appendix \ref{ap:jeans}, one can find analytic
formulae for the luminosity-weighted dispersions inside an aperture of radius
$A$, $\sigma^2_{\rm ap}(A)=\int_0^AI(R)\sigma^2(R)RdR/\int_0^AI(R)RdR$ for
systems composed of isotropic, radial or circular orbits in a logarithmic 
potential. Thus 
\[
\sigma_{\rm ap,iso}^2(A)=\frac{1}{3}v^2_c+ v^2_c\frac{A^3\int_A^\infty
  I(x)/x^2 dx}{3\int_0^A I(x)x dx} 
\]
\be
\sigma_{\rm ap,circ}^2(A)=\frac{1}{3}v^2_c - v^2_c\frac{A^3\int_A^\infty
  I(x)/x^2 dx}{\textstyle 6\int_0^A I(x)xdx}
\ee
\[
\sigma_{\rm ap,rad}^2(A)=\frac{1}{3}v^2_c+v^2_c\frac{A\int_A^\infty
  I(x) dx}{2\int_0^A I(x)xdx}  -v^2_c\frac{A^3\int_A^\infty
  I(x)/x^2 dx}{6\int_0^A I(x)xdx}.
\]
In Figure \ref{fig:aperture} the dependence of the luminosity-weighted
dispersions on the aperture size $A$ is shown for S\'{e}rsic laws with index
$n=1,4,12$.  The dispersions converge to the same value $v_c/\surd{3}$ at very
large aperture (as required by eq.\ \ref{eq:virial}), but strongly diverge at
small radii. For example for an $n=4$ S\'ersic model, the difference between
aperture dispersions $(\sigma_{\rm ap,rad}-\sigma_{\rm ap,circ})/v_c$ at
$A/R_e=0.5,1,2$ and 5 is 0.27, 0.17, 0.10 and 0.034; for comparison the
difference between the line-of-sight velocity dispersions $\sigma_{\rm
  iso}(R)-\sigma_{\rm rad}(R)$ evaluated at $R=R_s(n=4)$ is $0.034 v_c$.

We conclude that the aperture dispersion, as a measure of the circular speed,
is {\em more} sensitive to the anisotropy profile than the dispersion at a
single ``sweet spot'' radius $R_s$ unless the aperture is $\gtrsim 5R_e$. For
spatially resolved galaxies this implies that the dispersion at the sweet spot
is a more powerful tool for estimating the circular speed than the aperture
dispersion. Nevertheless the aperture dispersion is often the only
available quantity which can be used to estimate the circular speed \citep[e.g.,][]{2004NewA....9..329P}.

\subsection{Circular speed from velocity dispersion using local measurements}

\label{sec:local}

The methods outlined in the previous subsection require determining the
galaxy's effective radius and S\'ersic index, which is often difficult,
especially for the most massive and extended galaxies. Moreover, some galaxies
are not well described by a S\'ersic law. In some cases one can use equations
(\ref{eq:siso}), (\ref{eq:scirc}) or (\ref{eq:srad}) to predict the
line-of-sight velocity dispersion in terms of $v_c$ without first fitting to a
S\'ersic law; however, the use of these equations formally requires the
knowledge of $I(R)$ up to $R\to\infty$.

Let us now assume that we have an
observed surface-brightness profile $I(R)$, perhaps available only over a
limited range of radii, and we wish to estimate the circular speed $v_c$ from
measurements of the line-of-sight dispersion $\sigma(R)$ assuming that the
potential is isothermal. One can differentiate equations (\ref{eq:siso}),
(\ref{eq:scirc}), or 
(\ref{eq:srad}) with respect to $R$ to obtain a relation between the circular
speed and the {\em local} properties of $I(R)$ and $\sigma(R)$. Thus 
\[
\sigma_{\rm iso}^2(R)=v^2_c\frac{1}{1+\alpha+\gamma} 
\]
\be
\sigma_{\rm circ}^2(R)=v^2_c\frac{1}{2}\frac{\alpha}{1+\alpha+\gamma}
\label{eq:ag} 
\ee
\[
\sigma_{\rm rad}^2(R)=v^2_c \frac{1}{\left(\alpha+\gamma\right
  )^2+\delta-1}, 
\]
where 
\be
\alpha\equiv-\frac{d\log I(R)}{d\log R}, \ \ \gamma\equiv -\frac{d\log
  \sigma^2}{d\log R},\ \ \delta\equiv \frac{d^2\log[I(R)\sigma^2]}{d
  (\log R)^2}.
\label{eq:agd}
\ee 

\begin{figure}
\plotone{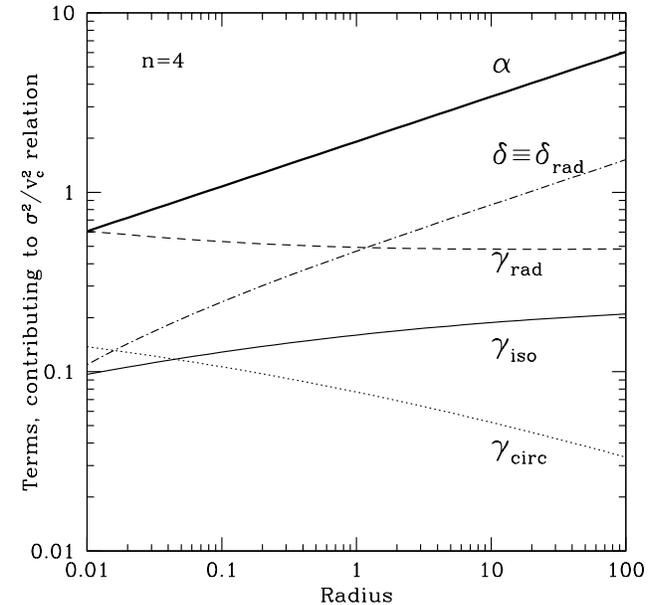}
\caption{Various terms contributing to the relation between the
  circular speed $v_c$ and the line-of-sight velocity
  dispersion. (see eq.\ \ref{eq:agd} for definitions). Each term is evaluated
  for an $n=4$ S\'ersic model. The $\gamma$ terms are calculated for
  isotropic, circular, and radial orbits; the $\delta$ term is
  calculated for radial orbits. The
  term $\alpha$ related to the slope of the surface-brightness
  profile (the upper curve) dominates.
\label{fig:terms}
}
\end{figure}

Although the terms $\gamma$ and $\delta$ can be
evaluated accurately if there is sufficiently good data on the dispersion
curve $\sigma(R)$, in a typical galaxy we expect these to be
subdominant compared to $\alpha$ and $\alpha^2$. This is illustrated
in Figure \ref{fig:terms} for a galaxy with S\'ersic index $n=4$.
If one neglects the subdominant terms then the expressions simplify further to
\[
\sigma_{\rm iso}^2(R)  =v^2_c\frac{1}{\alpha+1} 
\]
\be
\sigma_{\rm circ}^2(R) =v^2_c \frac{1}{2}\frac{\alpha}{\alpha+1}
\label{eq:pow} 
\ee
\[
\sigma_{\rm rad}^2(R)  =v^2_c \frac{1}{\alpha^2-1}. \\
\]
For pure radial orbits we can also use the expression which neglects
all derivatives of the line-of-sight velocity dispersion, but keeps
the second derivative of the surface brightness. Namely,
\be
\label{eq:raddelta} 
\sigma_{\rm rad}^2(R)  =v^2_c \frac{1}{\alpha^2+\delta_I-1},\\
\ee
where $\delta_I\equiv d^2\log[I(R)]/d (\log R)^2$.

\begin{figure}
\plotone{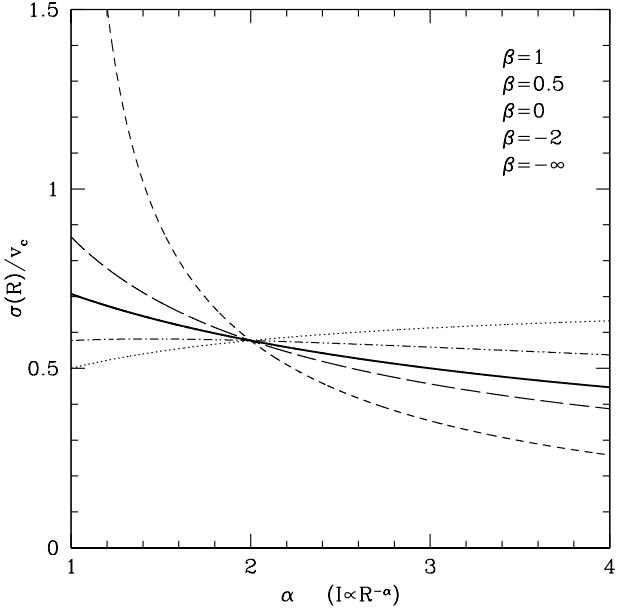}
\caption{Relation between the line-of-sight velocity dispersion and
  the circular speed for the case of a pure power law surface
  brightness distribution $I(R)\propto R^{-\alpha}$.
\label{fig:spower}}
\end{figure}

In the case of a power-law surface-brightness law $I(R)\propto
R^{-\alpha}$ in which the anisotropy parameter $\beta$ is independent of
radius the expression is (see also Figure \ref{fig:spower}):
\begin{eqnarray}
\sigma^2(R)=v^2_c\frac{1}{\alpha+1-2\beta}\left(
  1-\beta\frac{\alpha}{1+\alpha}\right). 
\label{eq:powb}
\end{eqnarray}
As pointed out by \cite{1993MNRAS.265..213G}, for $\alpha=2$ the anisotropy
parameter cancels out. In this case the relation between circular speed
$v_c$ and the 
line-of-sight velocity dispersion at a given radius $R$ solely depends
on the local slope of the surface-brightness profile (for $\beta={\rm const}$).

The above considerations suggest that one can get a reasonable estimate of the
circular speed using the observed surface brightness $I(R)$ and
line-of-sight dispersion $\sigma(R)$ profiles over a limited range of radii,
together with equations (\ref{eq:ag}). Similar to equations (\ref{eq:rsweet})
we can introduce different types of sweet spots (hereafter $R_2$) where pairs of orbital
models with observed quantities $\sigma(R)$, $\alpha$, $\gamma$
and $\delta$ correspond to the same circular speed\footnote{Note that these
  sweet spots are not the same as defined in \S\ref{sec:sweet}.}. In
particular
\be
\begin{array}{llll}
R_{2,ic}:&\alpha=2&&{\rm isotropic=circular} \\
R_{2,ir}:&\alpha+\gamma=2 &({\rm if~~}\delta=0)&{\rm isotropic=radial}. \\
\end{array}
\label{eq:rsloc}
\ee The expression for $R_{2,cr}$ (circular$\,=\,$radial) is more complicated (see
eq.\ \ref{eq:ag}) but based on the example of a S\'ersic profile we expect
that $R_{2,ir}<R_{2,cr}<R_{2,ic}$ so the range between the two radii in
equation (\ref{eq:rsloc}) should include $R_{2,cr}$. 
When both $\delta$ and $\gamma$ are small then all
three models intersect at the same point $R_2$ where $\alpha=2$ and
$\sigma_{\rm iso}(R_2)=\sigma_{\rm circ}(R_2)=\sigma_{\rm rad}(R_2)=
v_c/\surd{3}$. 

\begin{figure*}
\centering \leavevmode
\epsfxsize=1.9\columnwidth \epsfbox{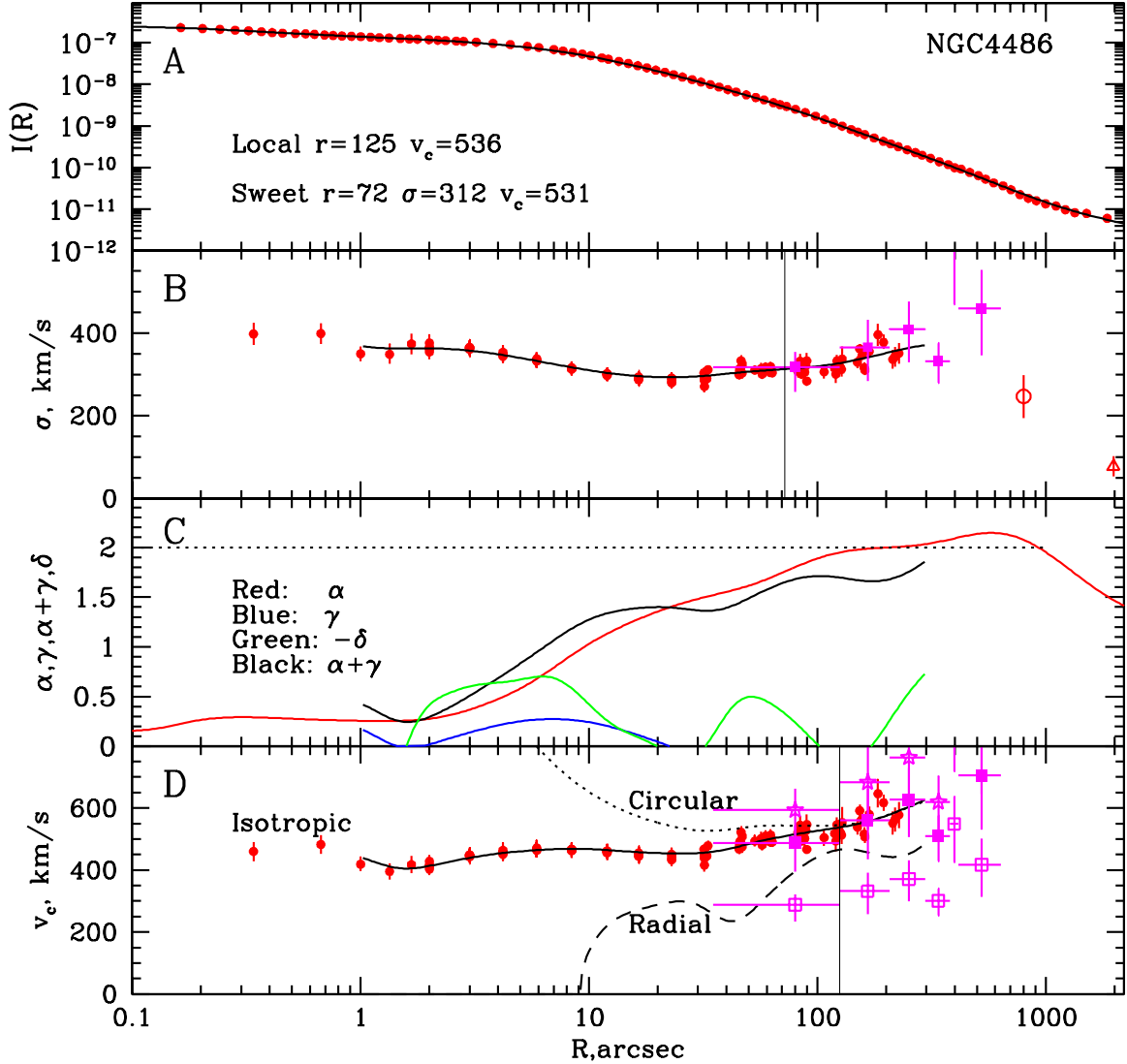}
\caption{Circular speed in M87 estimated from the line-of-sight
  velocity dispersion using equations (\ref{eq:ag}).  Shown in panel A
  is the observed surface-brightness profile in the V band
  from \protect\cite{2009ApJS..182..216K} ($I(R)=100^{-\mu/5}$, where $\mu$ is in
  magnitudes per arcsec$^2$). Shown in panel B is the
  line-of-sight dispersion profile from \protect\cite{doherty09}.
  Black solid lines in panels A and B show smooth fitting
  curves.  
  The magenta points with large error bars in
  panel B are the globular cluster data from
  \protect\cite{2001ApJ...559..828C}; 
  these are shown for reference but are not used in any fits. The
  last two data points at 800-2000$''$ are the PNe data from
  \protect\cite{2004ApJ...614L..33A}  (open circle) and  \protect\cite{doherty09} (open triangle).
  The fitting curves were used to compute $\alpha=-d\log
  I(R)/d\log R$, $\gamma=-d\log \sigma^2(R)/d\log R$, $\delta=d^2\log[I(R)\sigma^2]/d
  (\log R)^2$, and the sum
  $\alpha+\gamma$, which are shown in panel C by red, blue, green, and black
  lines.  Shown in panel D are the estimated values of the
  circular speed.  Red points in panel D correspond to the
 measured values of
  the line-of-sight velocity dispersion (shown in panel B), which were converted to
  the circular speed using equation (\ref{eq:ag}) for isotropic
  orbits and the fitted curve for $\alpha+\gamma$ in panel C.  The black solid, dotted and dashed curves are the
  result of applying equations (\ref{eq:ag}) for isotropic,
  circular and radial orbits respectively to the smoothed curves
  approximating the surface brightness $I(R)$ and the line-of-sight
  dispersion $\sigma(R)$.  The magenta points in panel D are the
  globular-cluster data converted to circular speed by approximating 
the surface density of globular clusters as a power law with
$\alpha\approx 1.35$ and
 using
  equation (\ref{eq:pow}) for circular (stars), isotropic (solid squares)
  and radial (open squares) models. Thin vertical lines in panels B and D show the radii used to estimate
the circular speed using the methods, described in \S\ref{sec:sweet}
and \S\ref{sec:local}, respectively.
\label{fig:m87}
}
\end{figure*}

\begin{figure*}
\plotwide{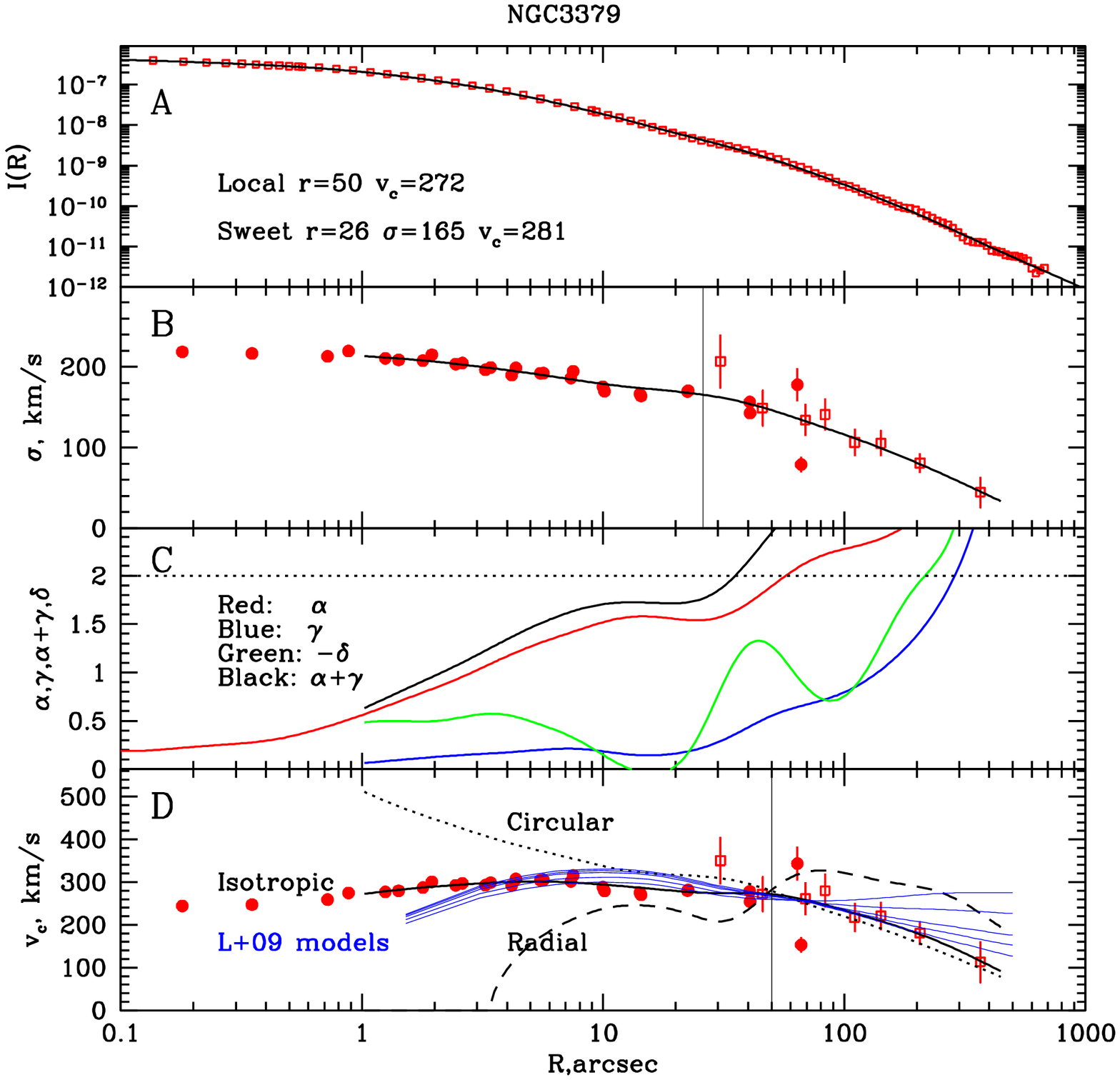}
\caption{The same as in Figure \ref{fig:m87} for NGC3379. Panel A shows the
  surface-brightness profile from wide-field B-band photometry by \protect\cite{1990AJ.....99.1813C}
 and 
HST V-band observations from \protect\cite{2000AJ....119.1157G}. The
velocity dispersion (panel B) is from \protect\cite{1999AJ....117..839S} and
\protect\cite{2000A&AS..144...53K}, supplemented by 
planetary nebula velocity dispersions (points shown as open squares with large error bars at
$R>30''$ in panel B, as calculated by \protect\citealt{2009MNRAS.395...76D} from data in \protect\citealt{2007ApJ...664..257D}).
Thin blue lines in panel D show
  the circular speeds from models A--E of  \protect\cite{2009MNRAS.395...76D}.
\label{fig:ngc3379}
}
\end{figure*}

We illustrate the application of this procedure to two well-studied
systems---M87 and NGC3379  (the first is in our X-ray sample but the second is not). Shown in Figures \ref{fig:m87} and
\ref{fig:ngc3379} are the observed surface brightness and line-of-sight
dispersion profiles (panels A and B) together with smooth curves
approximating their large-scale trends. 
For M87, the photometry is from \cite{2009ApJS..182..216K}.
The line-of-sight velocity dispersion is a combination of van der
Marel data \citep{1994MNRAS.270..271V} for the central regions, SAURON
\citep{2004MNRAS.352..721E} for the mid-radial regions and VIRUS-P
\citep{2008SPIE.7014E.231H} as presented in \cite{gebhardt2009b} for
the outermost stellar kinematics. \cite{2009ApJ...700.1690G} present an analysis
using the SAURON data, van der Marel data, and kinematics from
globular cluster kinematics.
For NGC3379
the photometry is from \cite{1990AJ.....99.1813C} and
\cite{2000AJ....119.1157G}, and the velocity dispersion is from 
\cite{1999AJ....117..839S} and \cite{2000A&AS..144...53K}.  We also use
planetary nebula velocity
dispersions at large distances where stellar velocity dispersions are not
available. For M87 we use data from \cite{2004ApJ...614L..33A} and
\cite{doherty09}, and for NGC3379 we use dispersions as calculated by
\cite{2009MNRAS.395...76D} from  data in \cite{2007ApJ...664..257D}. The use of these
data in our analysis is based on the empirical result that
the spatial distribution of
planetary nebulae tracks the spatial distribution of the stars
\cite{2009MNRAS.394.1249C}, but see the discussion below.

We then use the interpolated curves $I(R)$ and $\sigma(R)$ to
calculate the derivatives $\alpha$, $\gamma$ and $\delta$
defined in equation (\ref{eq:agd}). The values of $\alpha$, $\gamma$, $\delta$ and the
sum $\alpha+\gamma$ are shown in panel C with red, blue,
green and black lines respectively. The term $\delta$, which depends
on the second derivative of the surface brightness and the
line-of-sight velocity dispersion, is of course most sensitive to the
way the data are interpolated, but usually makes only a small contribution to the
estimate of circular speed. Some of the wiggles in $\delta$ seen in
panel C of Figures \ref{fig:m87} and \ref{fig:ngc3379}
are certainly spurious.

As we argued before there are two robust methods to estimate the circular
speed (for a logarithmic potential): (i) fit the photometry to a S\'ersic or
other model and determine the circular speed from the line-of-sight velocity
at the sweet spot $R_s$ (\S\ref{sec:methoda}); (ii) determine the circular speed from the
range of radii between points where $\alpha\approx2$ and $\gamma+\alpha\approx
2$ (\S\ref{sec:methodb}). At these radii the anisotropy of orbits does not affect the estimate of
the circular speed (all three black curves intersect in the vicinity of this
point).

M87 is fit well by a S\'ersic model with $n=6.5$, and the first method yields
$R_s=72$ arcsec and $v_c\simeq1.7\sigma(R_s)=530\kms$ (Tables \ref{tab:sample} and
\ref{tab:rs}). The second approach, based on radii $\sim125$ arcsec, yields
$v_c\approx 536\kms$ in good agreement. In principle one can extend
this analysis to larger radii using planetary nebulae (PNe), the data
points shown as red open symbols in Figures \ref{fig:m87} and
\ref{fig:ngc3379}. However, 
there are two problems in determining circular velocities from these outer
data points: (i) the
PNe dispersions are reported for $R\ge 800''$, leaving considerable freedom
in the interpolation of $\sigma(R)$ between $\sim 200''$ and $800''$; (ii)
at these large distances, the logarithmic surface brightness gradient
$\alpha$ applicable to the outer halo of M87 has some uncertainty, due
to both the uncertain details of the inferred truncation of the halo
\protect\cite{doherty09} and the possible 
contribution from the intracluster light 
(ICL) to the outer surface brightness profile.

NGC3379 is fit by a de Vaucouleurs model (a S\'ersic model with $n=4$) with
$R_e\simeq 50$ arcsec \citep{1990AJ.....99.1813C}, so the first method yields
$R_s\simeq 26$ arcsec and $v_c=1.7\sigma(R_s)\simeq 281\kms$. The second
approach, based on radii near $50$ arcsec, yields $v_c\simeq 272\kms$ in good
agreement. For NGC3379 the stellar and PNe dispersions overlap in radius, 
and no ICL has been found in the Leo group \cite{2003A&A...405..803C},
so we used the stellar surface brightness profile and the
combined stellar plus PNe
line-of-sight velocity dispersion profile (solid line in panel B of
Figure \ref{fig:ngc3379})
for the analysis.

\begin{figure*}
\centering \leavevmode
\epsfxsize=1.9\columnwidth \epsfbox{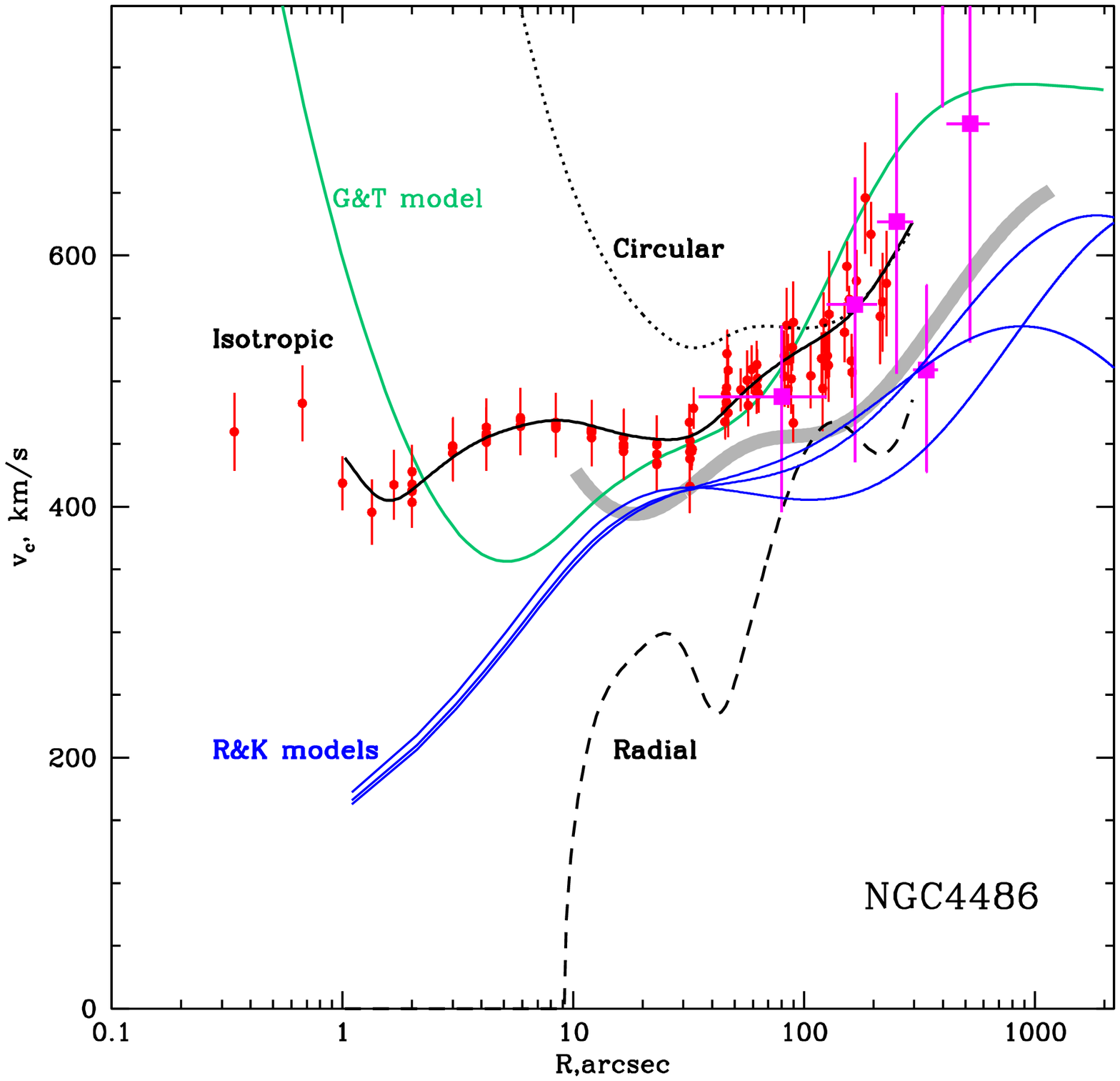}
\caption{Circular speed in M87. Enlarged version of panel D of
  Figure \ref{fig:m87}. Blue lines show the circular speeds from the
  detailed dynamical analysis of
  \protect\cite{2001ApJ...553..722R}---their models NFW1, NFW2,
  NFW3. The green line shows the best-fitting model of
  \protect\cite{2009ApJ...700.1690G}. The magenta points are the
  globular-cluster data converted to circular speed by approximating
  the surface density of globular clusters as a power law with
  $\alpha\approx 1.35$ and using equation (\ref{eq:pow}) for isotropic
  systems. The thick gray curve is the circular speed derived from heavily
  smoothed X-ray data.
\label{fig:m87vc}
}
\end{figure*}

In the bottom panel (D) of Figures \ref{fig:m87} and
\ref{fig:ngc3379} we show the estimated values of the circular
speed as a function of radius, for various assumptions about the velocity
anisotropy. Red points are derived from the values of the line-of-sight 
velocity dispersion, shown in panel B, which were converted to the
circular speed using
equation (\ref{eq:ag}) for isotropic orbits, i.e.,
\begin{eqnarray}
{v}_c(R)= \sigma_{opt}(R)\times \sqrt{1+\alpha+\gamma}.
\end{eqnarray}
The black solid, dotted, and dashed curves are the result of applying
equations (\ref{eq:ag}) for isotropic, circular and radial orbits respectively
to the smoothed curve approximating the line-of-sight dispersion profile
$\sigma(R)$ (black curve in panel B, see also Fig. \ref{fig:m87vc}). The agreement of the black solid curve
with the red points in panel D is a measure of the errors introduced by
smoothing. Our starting assumption was that the galaxy potential is
logarithmic at all radii (i.e., the circular speed is constant), so the
results in Figures \ref{fig:m87} and \ref{fig:ngc3379} are self-consistent
only if the circular speed shown in panel D is approximately independent of
radius. In M87 the solid black curve is rising at radii larger than
a few tens of arcseconds, while in NGC3379 the black curve is instead
declining with radius. This implies that either the
assumption of isotropy is incorrect---in which case the potential may still be
logarithmic, and the methods of \S\ref{sec:robust} will still give an accurate
assessment of the circular speed---or the assumption of constant circular
speed is incorrect. As discussed in \S\ref{sec:vcvar} if the orbits
are indeed isotropic then the black curves in panels D of Figures \ref{fig:m87} and
\ref{fig:ngc3379} are approximately correct. In case the anisotropy of
orbits is not known the estimates of the {\it local} circular speed
are still robust, but 
only in the vicinity of the radius where $\displaystyle -d\log
  I(R)/d\log R\approx 2$.

Thin blue lines in  Figure \ref{fig:m87vc} show the circular speeds
from published detailed dynamical analyses of M87 by 
\citealt{2001ApJ...553..722R}, their models NFW1, NFW2 and NFW3. The
green solid line in Figure \ref{fig:m87vc} shows the best fitting
model of \cite{2009ApJ...700.1690G}. The difference in the circular
speeds between \citealt{2001ApJ...553..722R} and
\cite{2009ApJ...700.1690G} can be traced largely to the difference in
the line-of-sight velocity dispersion data and to the absence of a central
black hole in the \citealt{2001ApJ...553..722R} model. Our estimates
of the circular speed, based on SAURON and VIRUS-P data of \cite{2009ApJ...700.1690G,gebhardt2009b} and the assumption of isotropic orbits,
agree reasonably well with \cite{2009ApJ...700.1690G} between
 20 and 200 arcsec. Compared to \cite{2001ApJ...553..722R} our
results (the solid black line) predict significantly larger circular
speed over the entire range of radii. This is not surprising, since
new measurements \citep{2009ApJ...700.1690G,gebhardt2009b} give larger
line-of-sight velocity dispersions.
They also disagree inside 10 arcsec but here the Romanowsky \& Kochanek
models exhibit substantial orbit anisotropy so we would not expect agreement
with our isotropic models. In addition, the Romanowsky \& Kochanek models do
not include a central black hole, which makes a significant contribution to
the circular speed inside 2--3 arcsec. Finally, the thick gray curve in
Figure \ref{fig:m87vc} is the circular speed derived from heavily
smoothed X-ray data. For this curve we used the potential profile obtained from
XMM-Newton data (see Fig.\ \ref{fig:pind}), which was smoothed using a
a Gaussian filter similar to the filter described in eq.\ \ref{eq:filter}
of Appendix \ref{ap:smooth} with $\Delta_R=0.5$. We differentiate
smoothed potential to obtain the circular speed $\displaystyle
v^2_c=r\,d\varphi/dr$. 
As discussed in \cite{2008MNRAS.388.1062C}, the
X-ray data agree well with the NFW2 model of
\cite{2001ApJ...553..722R}. This is also seen from the comparison of
circular speeds in Figure
\ref{fig:m87vc} - compare the thick gray line (X-rays) and the blue
line, corresponding to NFW2 model (the curve with the largest circular speed at $\sim 1000''$). 
The new optical data suggest substantially
higher circular speed and a larger contribution of the non-thermal
pressure is needed to bring the X-ray and optical data in agreement.
We also note that all methods suggest the increase of the circular
speed in M87 from $\sim$400-500$\kms$ inside central 2$'$ to $\sim$600-700$\kms$
outside 10$'$.

In NGC3379 our prescription for evaluation of the circular speed assuming
isotropic orbits (black solid line) is remarkably consistent with detailed
models (blue lines) between radii of 3 arcsec and 100 arcsec, even through
the circular speed in the detailed models is far from flat. We note
here that the models with the highest and lowest circular speeds at
400$''$ are formally ruled out by the likelihood analysis in
\cite{2009MNRAS.395...76D}; the preferred models are the middle three.

\begin{figure}
\plotone{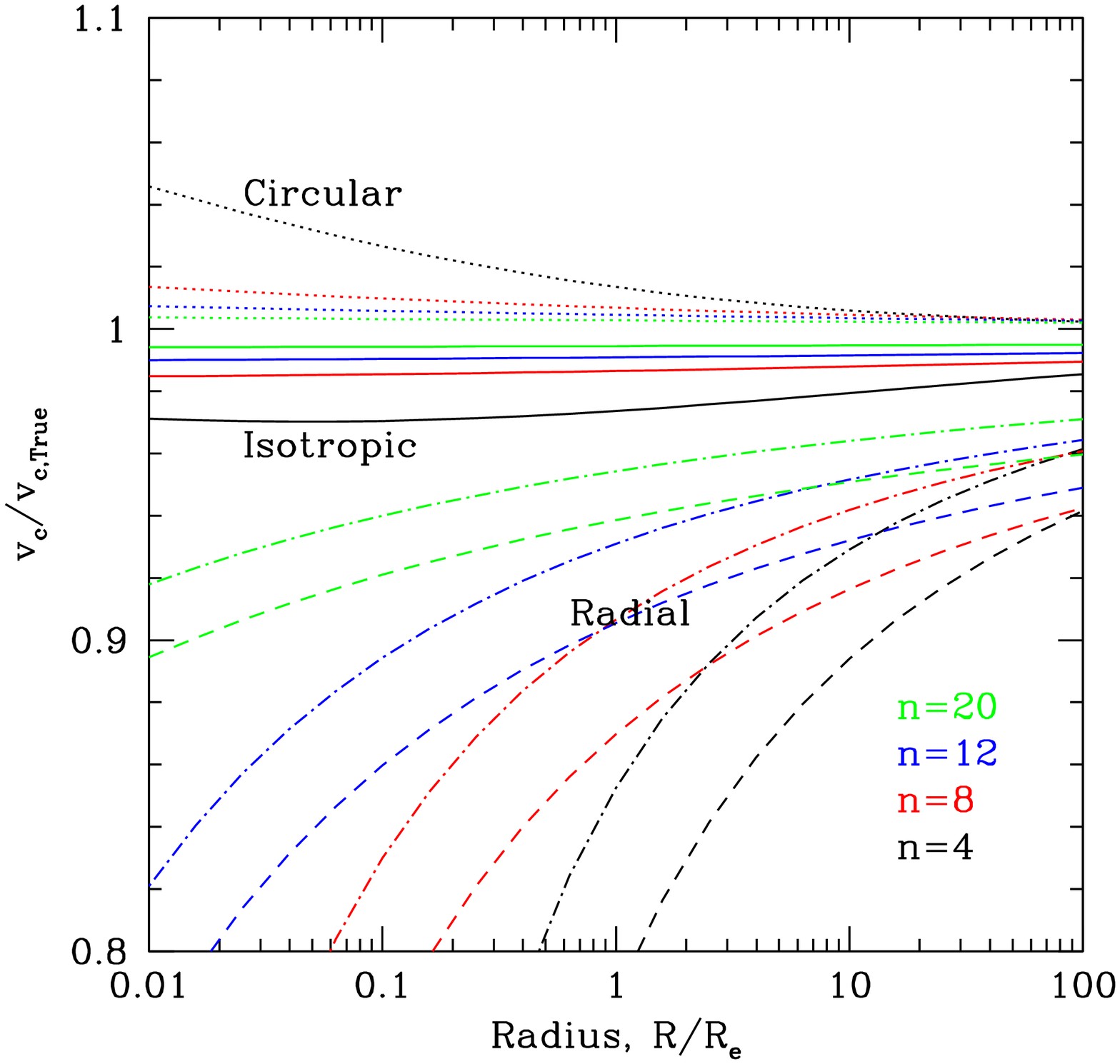}
\caption{Circular speed estimated from the line-of-sight velocity
  dispersion using approximate equation (\ref{eq:pow}) for isotropic (solid),
  circular (dotted) 
  and radial (dashed) orbits, instead of the exact equation
  (\ref{eq:ag}). The former equation uses a pure power law
  approximation to the surface brightness profile and completely
  ignores the terms $\gamma$ and $\delta$ (see
  \S\ref{sec:local}). For radial orbits more accurate values of the
  circular speed are obtained (dash-dotted lines) from another
  approximate equation 
  (\ref{eq:raddelta}). Colors denote S\'ersic models with $n=4,8,12$, and 20.
\label{fig:vcrec}}
\end{figure}

Alternatively one can neglect the contribution of $\gamma$ and
$\delta$ and use equation (\ref{eq:pow}) instead.  The accuracy of the recovered
value of $v_c$ is illustrated in Figure \ref{fig:vcrec} for several
values of the S\'ersic index $n$.
We used this simplified approach to plot the
data for M87 globular clusters from \cite{2001ApJ...559..828C}.  We plot the
line-of-sight velocity dispersions from Table 1 of C\^ot\'e et al.\ 
in Figure \ref{fig:m87} with magenta squares (panel B).
{}From Figure 12 of C\^ot\'e et al., we estimated the power law slope of
the surface density of globular clusters as $\alpha\approx 1.35$ and
used equation (\ref{eq:pow}) for isotropic, circular and radial models to
evaluate $v_c$ in panel D of Figure \ref{fig:m87}. As
one can see the range of values of $v_c$ spanned by these three models
overlaps reasonably well with the results based on the stellar velocity
dispersion, and the rise in $v_c$ for $r\gtrsim 100$
arcsec  appears to be mirrored in the
magenta points. However the rather shallow surface density of globular
clusters $I(R)\propto R^{-1.35}$ leads to relatively large
uncertainties in $v_c$.

\subsubsection{Circular speed varying with radius}
\label{sec:vcvar}

\begin{figure*}
\plottwo{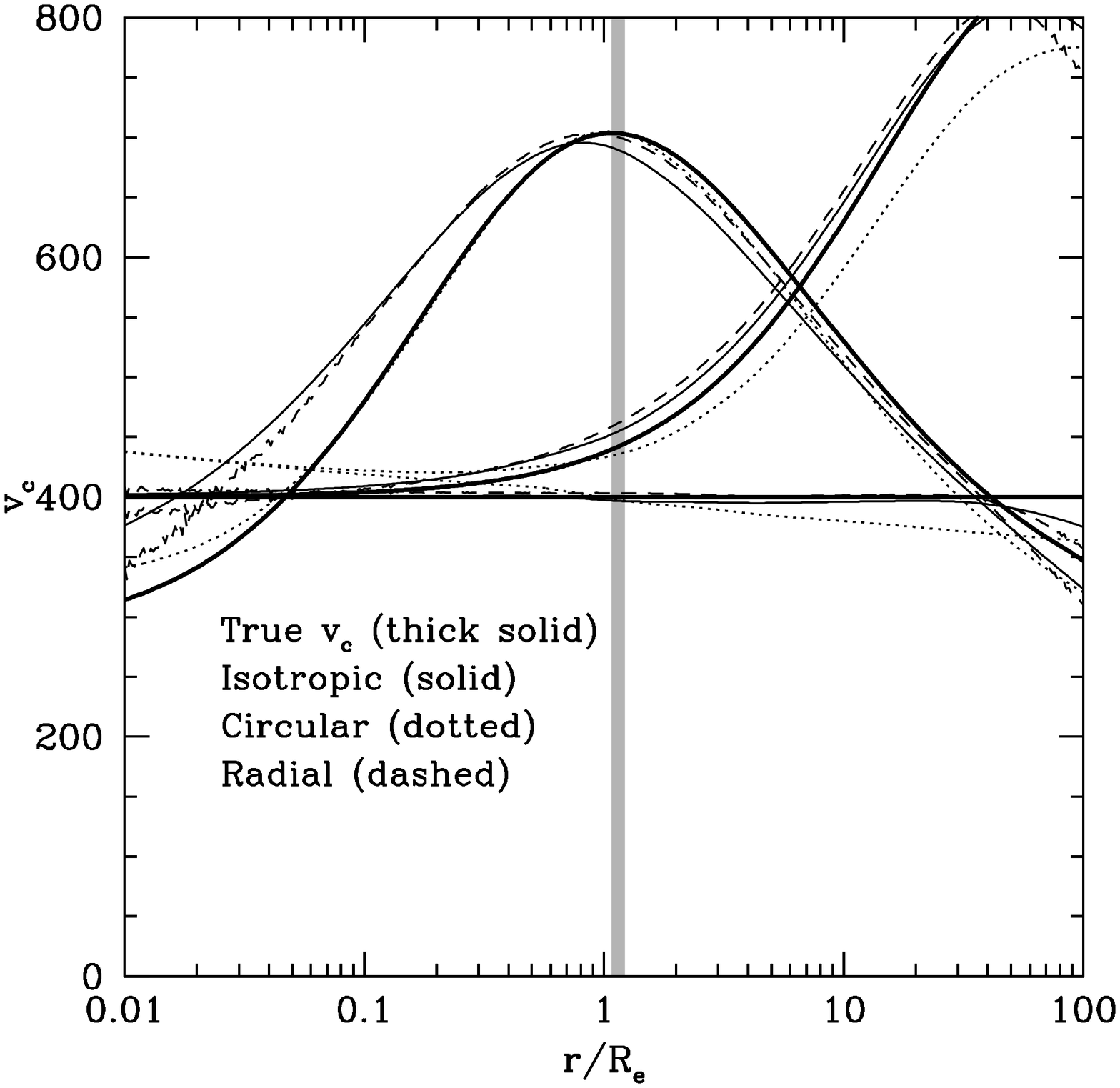}{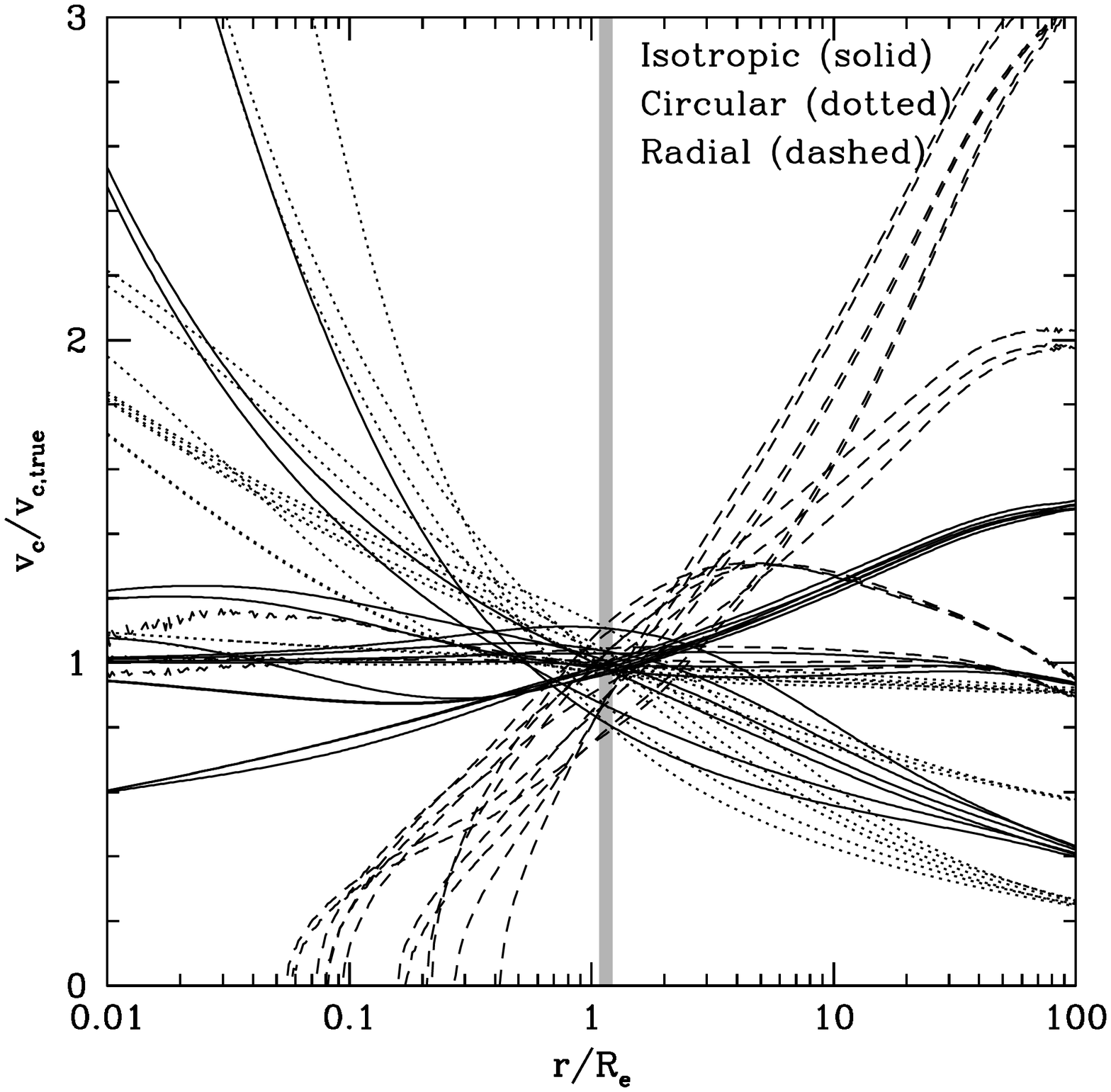}
\caption{{Left:} Recovery of the {\it non-constant circular speed}
  profile from the the line-of-sight velocity dispersion profiles
  using eq.\ \ref{eq:ag} when the {\it anisotropy parameter is
    constant and known}.  Three types of model $v_c(r)$ profiles are
  shown with thick solid lines: (i) $v_c(r)={\rm const}$, (ii)
  $v_c(r)$ steadily rising with radius and (iii) $v_c(r)$ having a
  maximum at $r \sim R_e$ and declining towards smaller and larger
  radii. For each of the $v_c(r)$ profiles the line-of-sight velocity
  dispersion $\sigma(R)$ was calculated for $\beta=$ 0, -9, and 1 and
  converted back to $v_c(r)$ using eq.\ (\ref{eq:ag}) for the isotropic
  (thin solid), circular (dotted) and radial (dashed) orbits. The
  vertical gray line shows the radius $R_{2}$ where $\displaystyle
  -\frac{d\log I(R)}{d\log R}=2$. The circular speed is recovered well
  even far away from $R_{2}$, if we know the anisotropy of orbit distribution.
  {Right:} Ratio of the recovered circular speed profile to the
  true circular speed profile when the true {\it $v_c(r)$ is
    non-constant} and the {\it (unknown) anisotropy parameter varies with radius}. The vertical gray line shows the radius
  $R_{2}$.  Without prior knowledge of the orbital anisotropy the circular
  velocity profile can not be recovered accurately, except in the
  vicinity of $R_{2}$ where the spread in the $v_c$ values (including
  models with variable $v_c$ and $\beta$) is small.
\label{fig:vcd}
}
\end{figure*}

Experiments in solving the Jeans equation for a galaxy with an $n=4$
S\'ersic surface-brightness profile have shown that the local
approximation works reasonably well at all radii even if the circular
speed is not constant, provided that the orbital anisotropy is known
and does not vary much with radius. This is illustrated in the left panel of Figure \ref{fig:vcd}. Three
types of model $v_c(r)$ profiles are shown in this figure with thick
solid lines: (i) $v_c(r)={\rm const}$, (ii) $v_c(r)$ steadily rising
with radius and (iii) $v_c(r)$ having a maximum at $r \sim R_e$ and
declining towards smaller and larger radii. For each of the $v_c(r)$
profiles we calculated the projected velocity dispersion $\sigma(R)$,
assuming constant anisotropy ($\beta=0,-9,1$) and converted
$\sigma(R)$ back to $v_c(r)$ using eq.\ \ref{eq:ag} for the
appropriate type of anisotropy. The recovered $v_c(r)$ curves are shown in
Figure \ref{fig:vcd} (left panel) with thin solid, dotted and dashed
lines for isotropic, circular and radial orbits respectively. The
vertical gray line shows the radius $R_{2}$ where
$\displaystyle -d\log I(R)/d\log R=2$, which is close to the
expected location of the region where the sensitivity of recovered $v_c(r)$ to
the orbital anisotropy is smallest. Clearly the circular speed is recovered well
even far away from $R_{2}$, if we can guess the anisotropy of the orbit distribution correctly.

If the anisotropy varies with
radius the error introduced in the circular speed is similar in
amplitude to that for the case of a logarithmic profile (Figure
\ref{fig:om}). Near the sweet point, defined according to equation
(\ref{eq:rsloc}), the value of $v_c$ is recovered to within 
  $\sim$10-20\%  even
if the circular speed is varying with radius. This is further illustrated in the right panel of Figure \ref{fig:vcd}. In
this figure we use the same three models of $v_c(r)$ profiles as in
the left panel and five different models for the orbital anisotropy: the constant anisotropies
$\beta=0,-9,1$, a model with orbits changing with radius from
isotropic to radial ($c=1$, $\beta_1=0$, $\beta_2=1$, $r_a=2$ in eq. \ref{eq:om}) and a model with orbits changing from isotropic
to circular ($c=1$, $\beta_1=0$, $\beta_2=-9$, $r_a=2$ in
eq. \ref{eq:om}). For each combination of $v_c(r)$
and $\beta(r)$ we calculate the projected velocity dispersion $\sigma(R)$
and convert it back to the circular speed eq.\ \ref{eq:ag} assuming
constant anisotropy $\beta=0, -\infty, 1$. This results in a total
$3\times 5 \times 3=45$ recovered circular speed profiles. The
ratio of the recovered profiles to the true circular speed profiles
is shown in Figure \ref{fig:vcd}. Not surprisingly without any prior knowledge of the orbital anisotropy the circular speed profile can not be
recovered accurately. However, in the vicinity of $R_{2}$ the spread
in the $v_c$ values (including models with variable $v_c$ and $\beta$)
is smallest: the ratio $\displaystyle v_c/v_{c,true}$ varies between
  0.78 and 1.12 for the whole set of models.

At the radii where $\alpha$ is far from 2 (either
smaller or larger) our method does not guarantee the recovery of
$v_c$ without a firm prior on the anisotropy parameter. At these radii
additional sources of information such as higher order velocity moments
($h_3$, $h_4$) and more
elaborate methods (like Schwarzschild's method) are needed.  One
example of a location where our method fails is the inner few arcsec in
M87 where a black hole of a few times $10^9~M_\odot$ dominates the mass. Because
the surface brightness is much flatter than $R^{-2}$ in this region
the allowed range for the circular speed $v_c$ (for anisotropy $\beta$ between
$-\infty$ and 1) is very wide (panel D of Fig. \ref{fig:m87}), easily large
enough to accommodate a black hole with a mass
ranging from 0 to $10^{10}~M_\odot$ or even larger\footnote{Note that
  for the innermost ($R\le 1''$) points in the line-of-sight velocity dispersion
  data (which are from \citealt{1994MNRAS.270..271V}) the seeing and the size
  of the slit can strongly affect the dispersion measurement. No
  attempts to correct for these effects were made in the present
  study}.

\subsection{Summary of suggested procedures for $v_c$ determination}

\subsubsection{For a galaxy with the surface-brightness profile well described
  by a S\'ersic law}

\label{sec:methoda}

Use Table \ref{tab:rs} (column 2) to find the ``sweet spot radius'' $R_s/R_e$
for a given $n$; use the line-of-sight velocity dispersion at this radius and
convert it to the circular speed using column 3 of Table \ref{tab:rs}; we call
this $v_{c,s}$ (see Table \ref{tab:sample}). Use column 4 from the same table
to crudely estimate the uncertainty in $v_c$.

\subsubsection{For a generic surface-brightness profile}

\label{sec:methodb}

Make a suitable interpolation of the surface-brightness profile (see,
e.g., Appendix \ref{ap:smooth}) and the line-of-sight velocity dispersion
profile. Calculate the slopes of these profiles (quantities $\alpha$,
$\gamma$, $\delta$, as defined in \S\ref{sec:local}). Convert
interpolated line-of-sight velocity dispersion into circular speed
using equations (\ref{eq:ag}). Locate the sweet spot where the three curves
are as close as possible (approximately corresponding to a range between radii
where $\alpha\approx2$ and $\alpha+\gamma\approx 2$) and use the
circular speed value predicted by the curves in the vicinity of this
radius. We call this $v_{c,l}$.  
When the curves do not intersect within the radial range well covered by the
data (or in which, at least, the extrapolation is reasonable) one can use
the radius where the range of circular speeds spanned by isotropic,  
circular and radial models is minimal. The difference between $v_c$
predicted by the three curves can be used to crudely characterize the
uncertainty.

\section{Results and discussion}
\label{sec:discussion}

Guided by the above considerations we compiled velocity dispersion
measurements from the literature for the galaxy sample in Table \ref{tab:sample}. 

\begin{itemize}

\item 
For the central velocity dispersion ($\sigma_{c}$) we rely on the
Hyperleda database. The corresponding circular speed was estimated as
$\displaystyle v_{c,c}=\sqrt{2}\sigma_{c}$, an approximation often used in the
literature but with little or no theoretical justification (column 9 in Table \ref{tab:sample}).

\item For the ``sweet spot'' estimates $R_s$ we use the values of the effective
radius and the S\'ersic index from \cite{1993MNRAS.265.1013C},
\cite{1994MNRAS.271..523D}, \cite{2005AJ....130.1502M}, and
\cite{2008MNRAS.385..667S}. The values of $R_s$ were
determined from the effective radius and S\'ersic index using Table
\ref{tab:rs}. The line-of-sight velocity 
dispersions at $R_s$ were taken from
\protect\cite{2000A&AS..144...53K},
\protect\cite{2008MNRAS.385..667S}, \protect\cite{doherty09} (and
references therein), and 
\protect\cite{2001ApJ...546..903D}.  The corresponding values $\sigma(R_s)$ are given in column
8 of the Table \ref{tab:sample}. In practice, for the selected
sample $R_s$ is always close to 0.5$R_e$ and the ratio
$\sigma_{iso}/v_c\simeq 0.6$. For one case when kinematic data do not
extend to $R_s$ (NGC4472) we used the outermost data points to evaluate
$\sigma$. Given the uncertainties in $R_e$, $n$ and the kinematics
data, the
accuracy of $\sigma$ is unlikely to be better than $\pm 5$--$10 \kms$.  In column 10 
we estimate the circular speed using equation  (\ref{eq:siso}) for
isotropic orbits (see also the entries for $\sigma_{\rm iso}$ in Table
\ref{tab:rs}).

\item For a second estimate of the circular speed we use the procedure
described in \S\ref{sec:local}: that is, we locate the region in which the logarithmic slopes
$\alpha$ and $\alpha+\gamma$ (eqs.\ \ref{eq:agd}) are as close as
possible to 2, find the
line-of-sight dispersion $\sigma$ at this radius, and then estimate
the circular speed using equation (\ref{eq:ag}) for an isotropic model, as
$\displaystyle v_{c,l}=\sigma\left (1+\alpha+\gamma\right )^{1/2}$. The values of
$v_{c,l}$ for each galaxy in our sample are given in column 11 of
Table  \ref{tab:sample}. The most uncertain is the value of $v_{c,l}$
for NGC4472, where the kinematic data of \cite{1994MNRAS.269..785B}
end at $R\simeq40''$, while the  S\'ersic approximation to the surface
brightness profile \citep{2009ApJS..182..216K} suggests that the 
intersection of curves occurs at $\sim200''$. 

\end{itemize}
\begin{figure}
\plotone{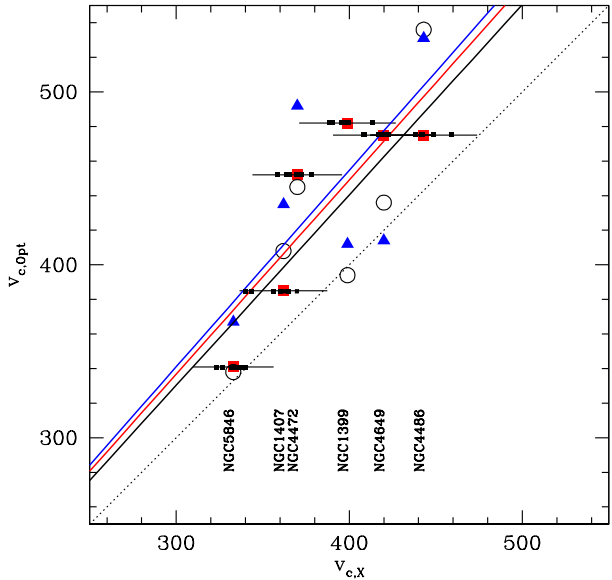}
\caption{Relation between $v_{c,X}$, the estimate of the circular speed from
  X-ray data, and three forms of $v_{c,opt}$, the estimate from the
  optical data. Red squares - $v_{c,c}$, blue
  triangles - $v_{c,s}$, black
  circles - $v_{c,l}$. The dashed line corresponds
  to $v_{c,0pt}=v_{c,X}$, and solid red, blue and black lines to the best
  fitting relations between $v_{c,X}$ and $v_{c,c}$, $v_{c,s}$ and
  $v_{c,l}$ respectively. Small black squares shows the
    values of $v_{c,X}$ obtained when changes are made to the analysis
    procedure (see \S\ref{sec:errx}). These changes include: letting
    the abundance of heavy 
    elements be a free parameter; adding a power law to account for
    potential contribution of LMXBs; varying lower or upper limits of
    the radial range used for $v_{c,X}$ calculations and separate
    calculations of the circular speed for the Northern and Southern
    parts of each galaxy. These
    different values of $v_{c,X}$ are plotted at fixed
    $v_{c,Opt}=v_{c,c}$ for each object. The black error bars show the
    crude estimate of the uncertainty in $v_{c,X}$ discussed in \S\ref{sec:errx}.
\label{fig:scor}
}
\end{figure}

Following the discussion in \S\ref{sec:sample} we compare the
circular speed derived from the X-ray analysis (Table
\ref{tab:sigmax}) with the three estimates of the circular speed
from the optical data as shown in Figure
\ref{fig:scor}.
 Assuming that
$v_{c,opt}=\eta\times v_{c,X}$ we derive the following best
fitting relations:
\begin{eqnarray}
\begin{array}{lll}
v_{c,c} &=1.12\times v_{c,X} & {\rm RMS}=0.07 \\
v_{c,s} &=1.14\times v_{c,X} & {\rm RMS}=0.11\\
v_{c,l} &=1.10\times v_{c,X} & {\rm RMS}=0.09,
\label{eq:scor} 
\end{array}
\end{eqnarray}
where root-mean-square deviations were evaluated as ${\rm RMS}=\left
[\sum_{i=1}^N \{v_{c,opt,i}/v_{c,X,i}- \eta \}^2/(N-1) \right]^{1/2}$,
where summation is over the objects in the sample, $N=6$. We note
here that the smallest scatter in the above relation is found
for the circular speed estimated from the central velocity dispersion
$v_{c,c}=\sqrt{2}\sigma_c$, rather than for more sophisticated
estimates, which were constructed to be least sensitive to unknown
orbit anisotropy. However, given the small size of our sample, no firm
conclusion on the quality of different proxies to the circular speed
is possible. 

 When five of the six objects are analyzed the value
  of $\eta$ varies from 1.07 to 1.13 for $v_{c,l}$ (compared to 1.10 for the
  full sample of six objects). The
  conclusion that $\eta > 1$ continues to holds. One can also estimate the
  significance of the statement that $\eta > 1$ from the fact that
  almost all of the points are in Figure \ref{fig:scor} are in the upper left part of the plot.
  Assuming that $\eta$ can equally probably be larger or smaller than
  unity the probability that 6 out of 6 points (for $v_{c,c}$) or 5 out
  of 6 (for $v_{c,s}$ and $v_{c,l}$) are found with $\eta >1$ is 0.016
  and 0.11 respectively.  

As discussed in \S\ref{sec:errx} changing the assumptions made in
  the analysis of X-ray data affects the value of $v_{c,X}$. This
  includes the effect of variable metalicity, contribution of LMXBs,
  dependence on the radial range used for evaluation of $v_{c,X}$ and
  independent calculations of the circular speed for the Northern and
  Southern parts of each galaxy. The values of $v_{c,X}$ obtained in
  each of the tests described in \S\ref{sec:errx} are shown with 
  small black squares in Figure \ref{fig:scor}. For clarity these
  values of $v_{c,X}$ are plotted at the same $v_{c,Opt}=v_{c,c}$. The
  black error bars show the crude estimate of the uncertainty in
  $v_{c,X}$ of $\sim$7\% obtained in \S\ref{sec:errx} under
  simplifying assumption that the contributions of different effects
  co-add quadratically.

As discussed in C08 the derived value of $v_{c,X}$ is sensitive to
non-thermal forms of pressure support\footnote{In C08
  instead of notation $\varphi=v_{c,X}^2\log r$  we
used $\varphi=2\sigma_X^2\log r$, where
$v_{c,X}\equiv \sqrt{2}\sigma_X$.}. This non-thermal support
can be parametrized by its impact on the relation between $v_{c,X}$
and the true circular speed $v_c$:
\begin{eqnarray}
v^2_c=v^2_{c,X} \left (1+ \frac{P_n}{P_g} \right ),
\end{eqnarray} 
where $P_n$ and $P_g$ are the non-thermal and gas thermal
pressures respectively. Thus the observed relation
(eq.\ \ref{eq:scor}) provides a measure of $P_n/P_g$: 
\begin{eqnarray}
\frac{P_n}{P_g}\approx \eta^2-1\approx 0.21-0.29,
\end{eqnarray}
depending on the adopted estimator of the optical velocity dispersion
in equation (\ref{eq:scor}). 

There are a number of obvious caveats (both on the X-ray and the optical sides)
associated with the above analysis. In particular there is
considerable uncertainty in the heavy metal abundance determination,
mentioned in \S\ref{sec:sample}, which may affect the estimates of
the potential at the level of a few per cent. Some of our galaxies sit in massive group/cluster halos; in
such cases the circular speed increases at large radii (e.g. M87 - see
Figure \ref{fig:m87vc}), so the approximation of the potential as a
logarithmic law characterized by a single circular speed may not be
sufficient for accurate comparison of the X-ray and optical data over
several decades of radius.
Minor deviations from
hydrostatic equilibrium, visible as wiggles in the potential profiles,
could also contribute to the scatter. Nevertheless we believe that our
main result---that the depth of the potential well derived from X-ray data
(from hot gas) is systematically
shallower (by few tens of percents) than the corresponding optical value (from stars)---is
robust. As mentioned in C08 these values are consistent with the current
paradigm of AGN controlled gas cooling/heating in the centers of
clusters and individual massive elliptical galaxies
\citep[e.g.,][]{2002MNRAS.332..729C}. The key
assumptions of this paradigm are that:
\begin{itemize}

\item Gas radiative cooling losses are balanced by the mechanical
  energy provided by a central black hole (AGN);

\item Dissipation of the mechanical energy is occurring on time
  scales comparable (within a factor of few) to the sound crossing
  time of the cooling region;

\item The ratio of the dissipation and cooling time scales (of the
  order of 0.1--0.2) sets the ratio of the non-thermal
    and thermal energy densities.
\end{itemize}

We finally note that the sample considered here is not statistically
complete and generalization of these results to all gas-rich
ellipticals must be done with caution.

\section{Conclusions}

\label{sec:conc}
Using non-parametric method we reconstructed gravitating
  potentials for a sample of six X-ray bright elliptical galaxies
  observed with Chandra and XMM-Newton. Their 
  gravitational potentials can be reasonably well
approximated by the isothermal (logarithmic) law $v_{c,X} \log r +b$
over the range of radii from $\sim$0.5 to $\sim$25 kpc, corresponding
to $\sim 0.05$--$3~R_e$. This result is in line with recent
lensing data which also suggest isothermality of the gravitational
potentials of the early-type galaxies and
  earlier results based on stellar kinematics and X-ray data. Many
galaxies in our sample are located at centers of massive groups/clusters and 
the X-ray data going beyond the optical extent of galaxies show the
steepening of the potential at large radii.

We suggest two new methods to derive an estimate of a galaxy's
circular speed if its potential is described by the isothermal
law. (i) For a spherical galaxy with a given S\'ersic index $n$, the
line-of-sight stellar velocity dispersion $\sigma_{opt}(R_s)$
evaluated at $R_s=R_s(n)\approx 0.5 R_e$ turns out to be relatively
insensitive to the anisotropy of stellar orbits. (ii) We suggest the
extension of this method for a generic surface-brightness profile
which allows one to estimate the circular speed through the ``local''
values of the line-of-sight velocity dispersion and the logarithmic
derivatives of the surface brightness and the velocity-dispersion
profiles.

Application of these methods to a sample of six massive elliptical
galaxies and the comparison to the results of X-ray analysis suggests
that on average 20\% of the gas pressure in these galaxies is
provided by non-thermal components (e.g., micro-turbulence or cosmic
rays). This result is in broad agreement with the current paradigm of
AGN controlled gas cooling/heating in the centers of clusters and
individual massive elliptical galaxies.

\section{Acknowledgments} 
We are grateful to Guinevere Kauffmann, Ben Metcalf, and Glenn van de Ven for useful
discussions. This work was supported
by the DFG grant CH389/3-2; NASA contracts and grants NAS8-38248, NAS8-01130,
NAS8-03060, and NNX08AH24G; the program ``Extended objects in the Universe'' of the
Division of Physical Sciences of the RAS; the Chandra Science Center;
the Smithsonian Institution; MPI f\"{u}r Astrophysik; MPI f\"{u}r
Extraterrestrische Physik and the Cluster
of Excellence ``Origin and Structure of the Universe''. ST acknowledges support from a Humboldt
Research Award.


\vspace{1cm}

\appendix

\section{Line-of-sight velocity dispersion in a logarithmic potential}

\label{ap:jeans}

The goal of this Appendix is to derive formulae for the line-of-sight velocity
dispersion profile $\sigma(R)$ for a spherical galaxy with surface-brightness
profile $I(R)$, assuming that the gravitational potential is logarithmic, 
\be
\phi(r)=v_c^2\log r + \hbox{const}.
\ee
The line-of-sight dispersion will depend on the shape of the
velocity-dispersion tensor, defined by its radial and tangential components
$\sigma_r^2(r)$ and $\sigma_\phi^2(r)=\sigma_\theta^2(r)$. We
examine three simple cases that should span the range of possible behaviors:
(i) isotropic orbits ($\sigma_r^2=\sigma_\phi^2=\sigma_\theta^2$);
(ii) radial orbits ($\sigma_\phi^2=\sigma_\theta^2=0$); (iii)
circular orbits ($\sigma_r^2=0$).

In a spherical system, the volume emissivity $j(r)$ and the surface
brightness $I(R)$ are related by
\begin{eqnarray}
I(R)&=&2\int_R^\infty\!\! {r\,dr\over\sqrt{r^2-R^2}}j(r), \nonumber \\
j(r)&=&-{1\over\pi}\int_r^\infty\!\! {dR\over\sqrt{R^2-r^2}}{dI\over dR}.
\label{eq:appone}
\end{eqnarray}

If the system is isotropic, the Jeans equation reads 
\be
{d\over dr}j\sigma_r^2=-j{d\phi\over dr}.
\ee
Specializing to the logarithmic potential and integrating
\be
j(r)\sigma_r^2(r)=v_c^2\int_r^\infty {du\over u}j(u).
\ee
The line-of-sight dispersion at projected radius $R$, $\sigma_{\rm iso}(R)$,
is given by
\begin{eqnarray}
I(R)\sigma_{\rm iso}^2(R)&=&2\int_R^\infty
{r\,dr\over\sqrt{r^2-R^2}}j(r)\sigma_r^2(r) \nonumber \\
&=&2v_c^2\int_R^\infty {r\,dr\over\sqrt{r^2-R^2}}\int_r^\infty {du\over
  u}j(u)\nonumber \\
&=& 2v_c^2\int_R^\infty {du\sqrt{u^2-R^2}\over u}j(u).
\end{eqnarray}
Replacing $j(u)$ from equation (\ref{eq:appone}) and exchanging the
order of integration,
\be
I(R)\sigma_{\rm iso}^2(R)=-{2v_c^2\over\pi}\int_R^\infty dx{dI\over
  dx}\int_R^x{\sqrt{u^2-R^2}\,du\over u\sqrt{x^2-u^2}}.
\ee
The inner integral is $\half\pi(1-R/x)$ so after integrating by parts
\be
I(R)\sigma_{\rm iso}^2(R)=Rv_c^2\int_R^\infty \frac{I(x)}{x^2}dx,
\ee
which is equation (\ref{eq:siso}). 

If the orbits are circular, a set of stars with random orientation at
radius $r$ and projected radius $R$ contributes a line-of-sight
dispersion $\half v_c^2R^2/r^2$. Thus the line-of-sight dispersion at
$R$ is given by 
\be
I(R)\sigma_{\rm circ}^2(R)=R^2v_c^2\int_R^\infty
{dr\over r\sqrt{r^2-R^2}}j(r).
\ee
Replacing $j(r)$ from equation (\ref{eq:appone}) and exchanging the
order of integration,
{\setlength\arraycolsep{2pt}
\begin{eqnarray}
I(R)\sigma_{\rm circ}^2(R)&=&-{R^2v_c^2\over\pi}\int_R^\infty dx{dI\over
  dx}\nonumber \\
 &&\quad\times \int_R^x{dr\over r\sqrt{r^2-R^2}\sqrt{x^2-r^2}}.
\end{eqnarray}}
The inner integral is $\half\pi/(Rx)$ so after integrating by parts
\begin{eqnarray}
\sigma_{\rm circ}^2(R)&=&\half v_c^2-{Rv_c^2\over 2I(R)}\int_R^\infty
  \frac{I(x)}{x^2}dx\nonumber \\
&=&\half v_c^2-\half \sigma_{\rm iso}^2(R),
\end{eqnarray}
which is equation (\ref{eq:scirc}). 

Finally, if the orbits are radial, the Jeans equation reads
\be
{d\over dr}r^2j\sigma_r^2=-jr^2{d\phi\over dr}.
\ee
Specializing to the logarithmic potential and integrating,
\be
j(r)\sigma_r^2(r)={v_c^2\over r^2}\int_r^\infty du\,uj(u).
\ee
If the orbits are radial, a set of stars at radius $r$ and projected 
radius $R$ contributes a line-of-sight
dispersion $\sigma_r^2(r)(r^2-R^2)/r^2$. The line-of-sight dispersion
at projected radius $R$, $\sigma_{\rm rad}(R)$, is then given by
{\setlength\arraycolsep{2pt}
\begin{eqnarray}
I(R)\sigma_{\rm rad}^2(R)&=&2\int_R^\infty
{dr\sqrt{r^2-R^2}\over r}j(r)\sigma_r^2(r) \nonumber \\
&=&2v_c^2\int_R^\infty {dr\sqrt{r^2-R^2}\over r^3}\int_r^\infty
du\,uj(u).
\end{eqnarray}}
Replacing $j(u)$ from equation (\ref{eq:appone}) and exchanging the
order of integration,
{\setlength\arraycolsep{2pt}
\begin{eqnarray}
I(R)\sigma_{\rm rad}^2(R)&=&-{2v_c^2\over\pi}\int_R^\infty
{dr\sqrt{r^2-R^2}\over r^3}\int_r^\infty dx\sqrt{x^2-r^2}{dI\over
  dx}\nonumber \\
&=&{2v_c^2\over\pi}\int_R^\infty
{dr\sqrt{r^2-R^2}\over r^3}\int_r^\infty {x\,dx\over
  \sqrt{x^2-r^2}}I(x) \nonumber \\ 
&=&{2v_c^2\over\pi}\int_R^\infty
dx\,xI(x)\int_R^x {dr\over r^3}{\sqrt{r^2-R^2}\over\sqrt{x^2-r^2}}.
\end{eqnarray}}
The inner integral is ${1\over4}\pi(x^2-R^2)/(Rx^3)$, so
\be
I(R)\sigma_{\rm rad}^2(R)={v_c^2\over 2R}\int_R^\infty
  I(x)(1-R^2/x^2)dx,
\ee
which is equation (\ref{eq:srad}). 

\section{Smoothing the surface-brightness profile}

\label{ap:smooth}
The procedure for evaluating the circular speed that is described in
\S\ref{sec:local} requires calculations of the derivatives of
the surface brightness $I(R)$ and the line-of-sight velocity
dispersion $\sigma(R)$. The derivative of the surface
brightness is expected to be the most important term (see
Fig.\ \ref{fig:terms}). The observed surface-brightness profiles
typically have enough data points with small errorbars to make the
selection of a simple analytic model difficult. To counter this
problem we used a simple procedure to smooth the data. Consider a set
of surface brightness measurements $I(R_i), i=1,\ldots,n$. Given
that we are interested primarily in logarithmic derivatives it makes
sense to use $f_i=\log R_i$ and $S_i=\log I(R_i)$ instead of $R$ and
$I$. We can then fit the linear relation between $f_i$ and $S_i$ as
$S_i=af_i+b$, where $a$ and $b$ are the free parameters of the fit. We
want to find ``local'' values of $a$ and $b$ in the vicinity of
a given radius $R_0$, which are based on a set of measurements not far
from $R_0$. We do that by choosing a weight function
\be
W(R_0,R)=\exp\left[-\frac{(\log R_0-\log R)^2}{2\Delta^2_R}\right], 
\label{eq:filter}
\ee 
where the parameter $\Delta_R$ controls the width of the weight function. The best
fitting parameters $a$ and $b$ (minimizing RMS deviation) for a given value of
$f_0\equiv \log R_0$ are given by 
\be
a(R_0)=\frac{\sum f_iW_iS_i\sum W_i-\sum W_iS_i\sum f_iW_i}{\sum f^2_iW_i 
  \sum W_i-(\sum_if_iW_i)^2}, 
\ee 
where $W_i=W(R_0,R_i)$, the sums are over $i=1,\ldots,n$, and
\be 
b(R_0)=\frac{\sum_iW_iS_i-a\sum_if_iW_i}{\sum_iW_i}.  
\ee 
The smoothed function $\tilde{I}(R)=e^{a(R)\log R +b(R)}$ is shown in panel A in
Figures \ref{fig:m87} and \ref{fig:ngc3379} with a black solid line. We used
$\Delta_R=0.3$ for both objects.

\label{lastpage}
\end{document}